\documentstyle[eqsecnum,epsfig,aps]{revtex}

\begin{document}
\draft
\preprint{FNT/T 98/03, PM/98-07}
\title{Rising bosonic electroweak virtual effects\\ 
at high energy $e^+e^-$ colliders}
\author{M. Beccaria}
\address{Dipartimento di Fisica, Universit\`a di 
Lecce and INFN, Sezione di Lecce\\
Via Arnesano, 73100 Lecce, Italy}
\author{G. Montagna and F. Piccinini}
\address{Dipartimento di Fisica Nucleare e Teorica, 
Universit\`a di Pavia and INFN, Sezione di Pavia\\
Via A.~Bassi 6, 27100 Pavia, Italy}
\author{F.M. Renard}
\address{Physique Math\'{e}matique et Th\'{e}orique, UMR 5825\\
Universit\'{e} Montpellier II, F-34095 Montpellier Cedex 5.}
\author{C. Verzegnassi}
\address{Dipartimento di Fisica, Universit\`a di 
Lecce and INFN, Sezione di Lecce\\
Via Arnesano, 73100 Lecce, Italy}
\date{\today}
\maketitle
\begin{abstract}
The virtual massive gauge boson effects on several observables in the
process of charged fermion pairs production at $e^+e^-$ colliders are
computed up to the TeV energy range in the Standard Model, working in
the ``$Z$-peak subtracted'' representation. It is shown that these
effects increase regularly with energy, approaching the typical ten
percent relative size. A careful numerical evaluation is proposed,
where the conditions dictated by gauge invariance are consistently
implemented.
\end{abstract}


\section{Introduction}
\label{sec:intro} 

The accurate computation of virtual electroweak effects has appeared
to be a fundamental theoretical contribution for a correct
interpretation of the experimental results obtained in several years of
running at the $Z$ resonance at LEP1 and SLC~\cite{rev}. Leaving aside the
remarkable degree of accuracy to which the Standard Model (SM) has been
tested at the one loop level, the prediction for the value of the top
mass (spectacularly confirmed by its experimental discovery) and the
subsequent limits on the ``low'' value of the Higgs mass undoubtedly
represent an illustration of how relevant a rigorous calculation of
virtual effects may be.

From a purely technical point of view, a calculation of electroweak
(i.e. of not conventional QED type) virtual effects appears to be
somehow simplified on top of the $Z$ resonance, since for trivial
reasons the non resonant terms can be evidently neglected. This
eliminates the full set of ``box'' diagrams, thus reducing the
theoretical task to that of computing self-energies and vertices. In
the self-energies, the dominant effect is provided by light fermion
pairs that determine the running of $\alpha^{(QED)}$ up to the $Z$
mass~\cite{alfa}, and by the heavy top that shifts from $1$ the value of
the $\rho$ parameter~\cite{Veltman}.

A welcome feature of these fermionic contributions is
that they are separately gauge independent. This   
is also true for what concerns the remaining ``large'' $\simeq m^2_t$
 virtual effect provided by the $Zb\bar b$ vertex~\cite{Zbb}. 
 Therefore, the full
set of ``relevant'' virtual electroweak contributions at the $Z$ peak
appears to be relatively ``simple'', the main role being played by either
the fermionic content of self-energies or by a gauge independent subset
of the bosonic effects.

As soon as one moves away from the special LEP1/SLC configuration, the
expectation is that the previous simple features might disappear. In
particular, the role of boxes (where $s$-channel $W$ exchange is
intrinsically not gauge independent), could become much more relevant.
Should this be the case, one extra immediate complication would arise
since the gauge dependence e.g. of $W$ bosons must be cancelled, in the
contributions to physical observables, by that of suitable combinations
of corresponding self-energies and vertices where virtual $W$s are
exchanged.
Clearly, only the sum of such special box, self-energies and vertices
combinations can acquire a physical meaning, and must be computed as
carefully as possible.

To make the previous discussion more quantitative, we can now provide
one numerical example, i.e. the calculation of the box contribution to
cross sections and asymmetries in the LEP2 energy range. Working in the
t'Hooft $\xi=1$ gauge, one sees~\cite{B200} that, at the largest
considered energy $\sqrt{s} \simeq 200$~GeV, the related effect can become
quite respectable (a relative one percent in the muon cross section and
a two-three percent in the hadronic cross section). Since this effect is
apparently increasing with energy, a question that naturally arises at
this stage is that of whether this feature will remain valid, or
stronger, at the future linear $e^+e^-$ colliders, whose energies are
expected to lie in the $500$~GeV range (or more~\cite{LC}). In
particular, the problem that one should face is that of computing
correctly the relevant gauge-invariant combinations of self-energies,
vertices and boxes. In fact, if their size became ``quite'' respectable,
the extra problem of the calculation of the initial-state QED
radiation corrections should also be considered in a self-consistent
way i.e. including all the various terms in the convolution process that
is usually involved~\cite{conv}.

The aim of this paper is precisely that of computing, in a fully gauge
invariant way, the contribution of bosonic effects to physical
observables of $e^+e^-$ annihilation into charged fermion pairs. This
will be done in an energy range from $200$~GeV to $1$~TeV (although we
could increase the upper limit if requested), working systematically in
our chosen subtracted theoretical framework. As we shall see, the
numerical effect generally increases with energy, approaching a typical
ten percent size at the TeV scale. This will be fully discussed in
the following Sec.\ \ref{sec:bos} and Sec.\ \ref{sec:pred}, where the 
numerical calculation of a few relevant observables will be performed 
without taking photon
emission into account. The large size of the effect will then motivate a
rigorous calculation of the initial-state QED radiation corrections; this 
will be treated in Sec.\ \ref{sec:qed}. A final brief discussion will then be
given in the concluding Sec.\ \ref{sec:concl}. An appendix contains explicit
formulae that are involved in the procedure but not essential for a
qualitative understanding of the main results.

\section{Bosonic virtual effects in the ``$Z$-peak subtracted''
representation}
\label{sec:bos}

It has been recently shown~\cite{R7,R8} that the process of $e^+e^-$
annihilation into fermion pairs, at squared center of mass (c.m.) 
energy $q^2$ larger than $M^2_Z$, can be conveniently described 
using an effective
parameterization, strictly valid at the one loop level, where only
subtracted quantities appear as ``electroweak corrections''. In
particular, the contributions to all observables generated by the $Z$
boson exchange are systematically subtracted on top of the $Z$ peak, at
$q^2=M^2_Z$, in full analogy with the photon contributions that are
subtracted ``on top'' of the photon peak at $q^2=0$. As a consequence of
this procedure, the theoretical input in the various $Z$ contributions
will contain quantities (partial $Z$ widths and asymmetries) measured
at the $Z$ resonance. These input parameters will replace the commonly
used Fermi coupling $G_{\mu}$, and we devote the interested reader to
Refs.\ \cite{R7,R8} for an exhaustive discussion of the details of the
approach. The final result is that, in the complete expression of a
general observable of the process $e^+e^-\to f\bar f$, there will be
some ``leading'' terms containing $\alpha_{QED}$, $M_Z$ and $Z$ peak
observables and four independent one loop subtracted corrections that
contain all the residual physical information and are defined as:

\begin{equation}
\widetilde{\Delta}_{\alpha,
ef}(q^2,\theta)=\widetilde{F}_{\gamma\gamma,ef}(0,\theta)-
\widetilde{F}_{\gamma\gamma,ef}(q^2,\theta)
\label{E1}
\end{equation}

\begin{equation}
R_{ef}(q^2,\theta)= I_{Z,ef}(q^2,\theta)-I_{Z,ef}(M^2_Z,\theta)
\label{E2}
\end{equation}

\begin{equation}
V^{\gamma Z}_{ef}(q^2,\theta)=
\frac{\widetilde{A}_{\gamma Z,ef}(q^2,\theta)}{q^2}-
\frac{\widetilde{A}_{\gamma Z,ef}(M^2_Z,\theta)}{M_Z^2}
\label{E3}
\end{equation}

\begin{equation}
V^{Z\gamma}_{ef}(q^2,\theta)=
\frac{\widetilde{A}_{Z\gamma,ef}(q^2,\theta)}{q^2}-
\frac{\widetilde{A}_{Z\gamma,ef}(M^2_Z,\theta)}{M_Z^2}
\label{E4}
\end{equation}

In the previous expressions, we have used the following definitions:
\widetext

\begin{equation}
I_{Z,ef}(q^2,\theta)= {q^2\over
q^2-M^2_Z}[\widetilde{F}_{ZZ,ef}(q^2,\theta)
-\widetilde{F}_{ZZ,ef}(M^2_Z,\theta)]
\label{E5}
\end{equation}

\begin{equation}
\widetilde{A}_{ZZ,ef}(q^2,\theta)=
\widetilde{A}_{ZZ,ef}(0,\theta)+q^2\widetilde{F}_{ZZ,ef}(q^2,\theta)
\label{E9}
\end{equation}

\begin{equation}
\widetilde{A}_{ZZ,ef} (q^{2}, \theta) = 
A_{ZZ} (q^{2}) - (q^2-M^2_{Z})[
(\Gamma_{\mu,e}^{(Z)} , v_{\mu,e}^{(Z)}) +
(\Gamma_{\mu,f}^{(Z)} , v_{\mu,f}^{(Z)})+
(q^2-M_Z^2) A^{(Box)}_ {ZZ, ef} (q^{2}, \theta)]
\label{E10}
\end{equation}

\begin{equation}
\tilde{F}_{\gamma\gamma,ef} (q^{2}, \theta) = 
F_{\gamma\gamma,ef} (q^{2}) -
(\Gamma_{\mu,e}^{(\gamma)} , v_{\mu,e}^{(\gamma)}) -
(\Gamma_{\mu,f}^{(\gamma)} , v_{\mu,f}^{(\gamma)})-
q^2 A^{(Box)}_ {\gamma\gamma, ef} (q^{2}, \theta)
\label{E6}
\end{equation}

\begin{eqnarray}  
{\widetilde{A}_{\gamma Z,ef} (q^{2}, 
\theta)\over q^2}
&=& {A_{\gamma Z} (q^{2})\over q^2}
 - ({q^2-M^2_{Z}\over q^2})
(\Gamma_{\mu,f}^{(\gamma)} , v_{\mu,f}^{(Z)})\nonumber\\
 &&-(\Gamma_{\mu,e}^{(Z)} , v_{\mu,e}^{(\gamma)})-(q^2-M^2_{Z})
A^{(Box)}_ {\gamma Z, ef} (q^{2}, \theta)
\label{E7}
\end{eqnarray}

\begin{eqnarray}  
 {\widetilde{A}_{Z\gamma,ef} (q^{2}, \theta)\over q^2}
&=& {A_{\gamma Z} (q^{2})\over q^2} - {q^2-M^2_{Z}\over q^2}
(\Gamma_{\mu,e}^{(\gamma)} , v_{\mu,e}^{(Z)})\nonumber\\
 &&-(\Gamma_{\mu,f}^{(Z)} , v_{\mu,f}^{(\gamma)})-(q^2-M^2_{Z})
A^{(Box)}_ {Z\gamma, ef} (q^{2}, \theta)
\label{E8}
\end{eqnarray}

The quantities $A_{ij}(q^2)= A_{ij}(0)+q^2 F_{ij}(q^2)$ ($i, j=\gamma, Z$) are 
the conventional transverse $\gamma,Z$ self-energies. 
$A^{(Box)}_{\gamma\gamma,\gamma Z, Z \gamma, ZZ, ef}(q^2,\theta)$ 
are the projections on the photon and $Z$
Lorentz structures of the box contributions to the scattering amplitude 
${\cal A}_{ef}$
and the various brackets ($\Gamma_{\mu},v_{\mu}$) are the projections
of the vertices on the different Lorentz structures to which
$\widetilde{A}_{\gamma\gamma}$,  $\widetilde{A}_{ZZ}$, 
$\widetilde{A}_{\gamma Z}$, $\widetilde{A}_{Z \gamma}$ belong.
In our notations $A^{(Box)}_{\gamma\gamma}$ is the 
component of the scattering
amplitude at one loop that appears in the form 
$v^{(\gamma)}_{\mu, e} A^{(Box)}_{\gamma\gamma} v^{(\gamma), \mu}_{f}$
where $v^{(\gamma),\mu}_{e,f} \equiv -|e_0|Q_{e,f}  \gamma^{\mu}$
is what we call the photon Lorentz structure 
and analogous definitions are obtained for  $A^{(Box)}_{ZZ}$, 
$A^{(Box)}_{\gamma Z}$, $A^{(Box)}_{Z \gamma}$
with the $Z$ Lorentz structure defined as 
$v^{(Z),\mu}_{e,f} \equiv
-{|e_0|\over2s_0c_0}\gamma^{\mu}(g^0_{V,e,f}-g^0_{A,e,f}\gamma^5)$. 
We do not insist
here on these technical points that are, we repeat, fully discussed in
previous references~\cite{R7,R8,clean,tc}. Our notations are almost
rigorously following the conventions of Degrassi and Sirlin~\cite{DG1,DG2}, 
and our approach is often motivated by their previous
observations. In particular, one can immediately understand 
that all the four quantities listed 
in Eqs.\ (\ref{E1})-(\ref{E4}) are separately gauge-independent, 
since they
contribute different Lorentz structures of the process. As a
consequence of this fact, whenever computing contributions e.g. from
virtual $W$s exchanges, that are intrinsically gauge dependent, it
will be essential for a clean interpretation of the result to stick the
various effects in the proper previous combinations.

\begin{figure}[hbt]
\begin{center}
\epsfig{file=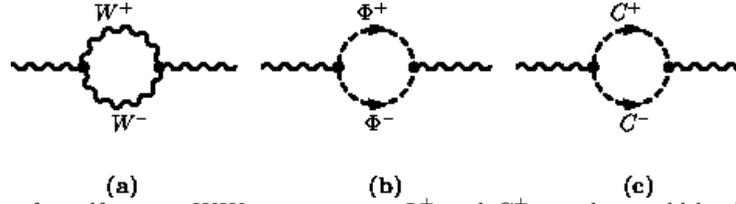,height=2.6truecm}
\caption{Feynman diagrams for self-energy $WW$ components. 
$\Phi^{\pm}$ and $C^{\pm}$ are the would-be Goldstone bosons and 
ghosts, respectively. }
\label{fig1}
\end{center}
\end{figure}

To illustrate the previous qualitative statements with a concrete
example, we start from the calculation of the charged boson
contributions to the generalized photonic correction
$\widetilde{\Delta}_{\alpha, ef}(q^2,\theta)$ of Eq.\ (\ref{E1}). Since this
quantity is by construction gauge-independent, we have performed the
calculation in the familiar t'Hooft gauge. This means that the
contributions from the unphysical charged would-be Goldstone bosons
must be incorporated and added to those coming from the charged $W$s
(and ghosts). We shall follow the order given by the various
definitions and compute the transverse self-energies terms first. In
the case of $\widetilde{\Delta}_{\alpha, ef}$, these correspond to the
Feynman diagrams shown in Fig.\ 1. A straightforward calculation leads to
the result (only the sum of all separate terms of Fig.\ 1 is given):

\begin{eqnarray}
&&\widetilde{\Delta}^{s.e.~(WW)}_{\alpha, e\mu}(q^2)=
F^{1a+1b+1c}_{\gamma\gamma}(0)-F^{1a+1b+1c}_{\gamma\gamma}(q^2)=\nonumber\\
&&-\left({5\alpha\over4\pi}\right)
\{\int^1_0 \!dx ~[1-{12\over5}x(1-x)]\ln|1-{q^2\over
M^2_W}x(1-x)| \nonumber\\
&&+{4\over15}+{8M^2_W\over5q^2}\int^1_0\! dx ~\ln|1-{q^2\over
M^2_W}x(1-x)|\}
\label{E11}
\end{eqnarray}

\vfil\eject

\begin{figure}
\begin{center}
\epsfig{file=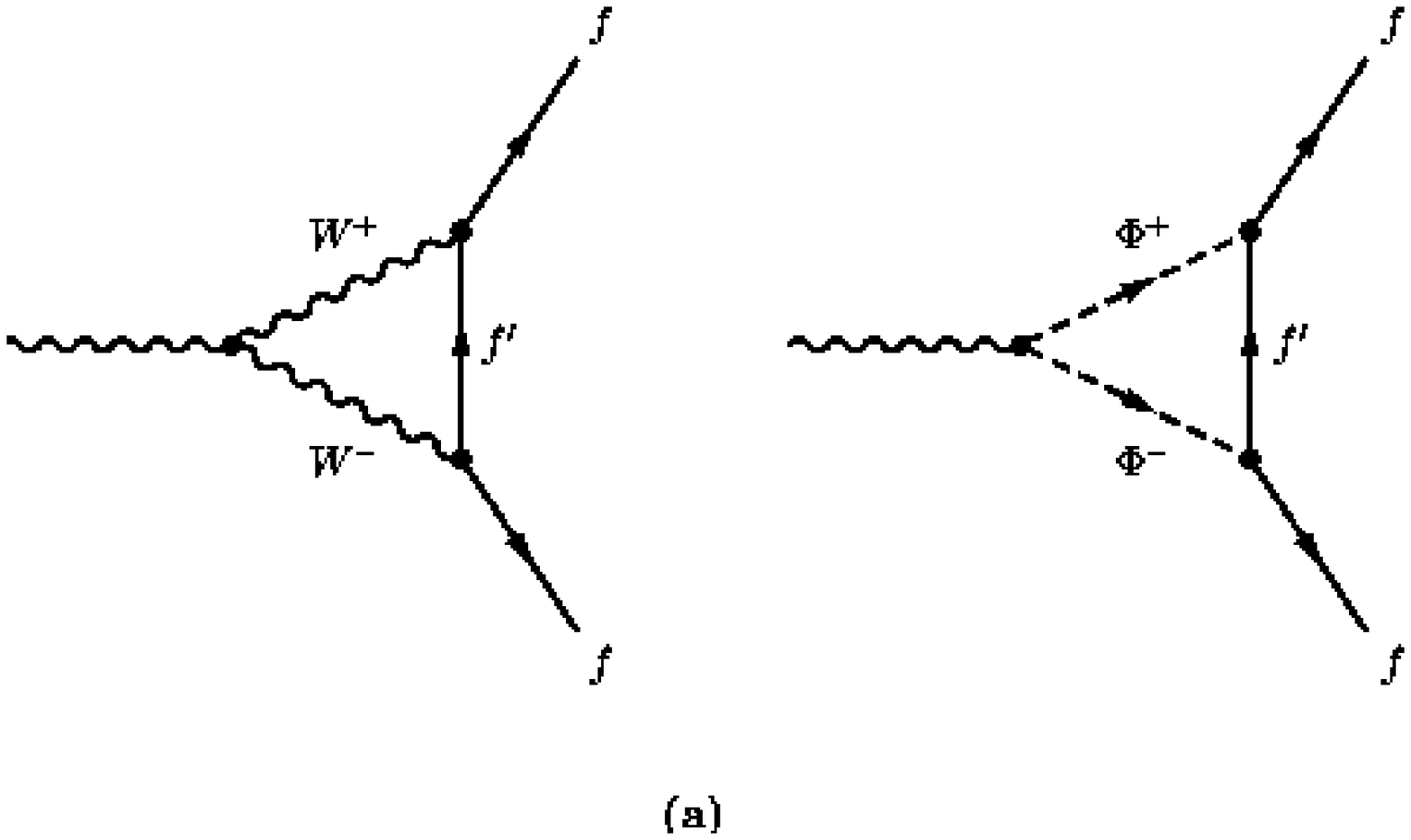,height=3.truecm}
\epsfig{file=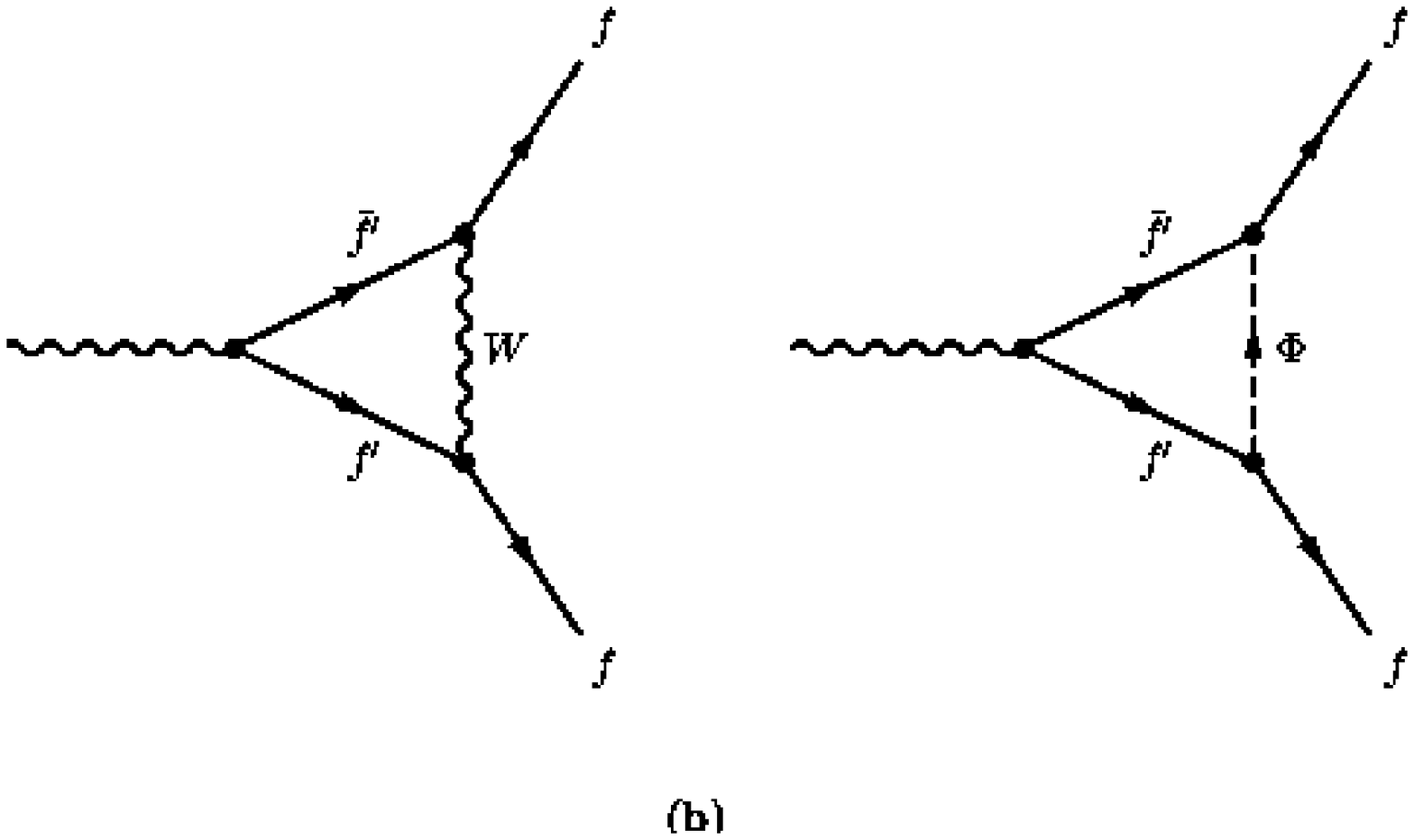,height=3.truecm}
\epsfig{file=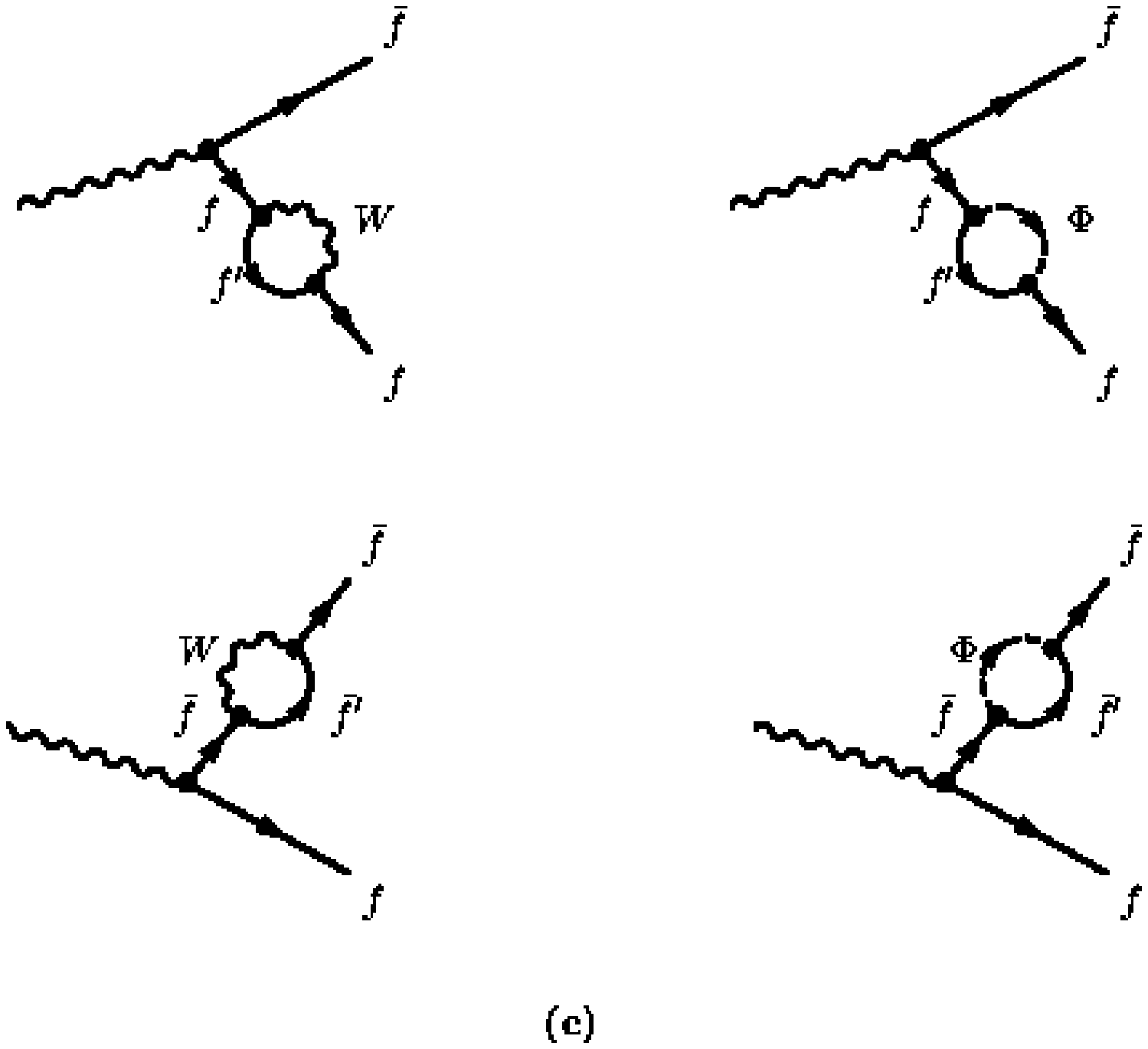,height=3.truecm}
\caption{Feynman diagrams for $WW$ vertex component 
(Fig.~2a), for single $W$ vertex (Fig.~2b)
 and for fermion self-energy 
diagrams (Fig.~2c).}
\end{center}
\label{fig2}
\end{figure}

The previous self-energy component 
$\widetilde{\Delta}^{s.e.~(WW)}_{\alpha, ef}$
is universal i.e. independent of the final fermion features. This is
not true for the remaining vertex and box contributions. It is
convenient therefore to consider the various cases separately by
treating final muons (or taus), final $u$ quarks and final $d$ quarks
states in succession. Starting with final muons, we have first
to consider the diagram involving $W$s in Fig.\ 2a  
(we are assuming massless final fermions, so that their
couplings with would-be Goldstone $\Phi$ bosons vanish). There is also
no photon-neutrino coupling so no diagram of the type of Fig.\ 2b.
In principle one should finally add the 
fermion self-energy diagrams of Fig.\ 2c. However their contribution
is $q^2$-independent and disappears in the subtraction. This is also
the case for any type of tadpole diagram, that we therefore never
discuss here.
This leads to the following expression of the (two $W$s) contribution to
the generalized vertex $\Gamma_{\mu,\gamma}(q^2)$, to be conventionally
defined as $\Gamma^{(WW)}_{\mu,\gamma}(q^2)$:
\widetext

\begin{eqnarray}
&&\Gamma^{(WW)(2a+2b)}_{\mu,\gamma}(q^2)=-\left({e^3\over
64\pi^2s^2}\right)\gamma_{\mu}(1-\gamma^5)\{-2({1\over\epsilon}-\bar
\gamma)+{3\over2}-\ln M^2_W\nonumber\\
&&+2\int\!\int\! dx_1dx_2~[~3~\ln |M^2_W(x_1+x_2)-q^2x_1x_2|
+{q^2(x_1+x_2-x_1x_2)\over (M^2_W(x_1+x_2)-q^2x_1x_2)}]\}
\label{E12}
\end{eqnarray}
where $\epsilon = (4-n)/2$ and $\bar\gamma = \ln (4 \pi ) + \gamma$, $\gamma$ 
being the Eulero-Mascheroni constant.

Projecting this vertex on the photon Lorentz structure, according to
our procedure, gives then

\widetext
\begin{equation}
\widetilde{\Delta}^{vert.~(WW)}_{\alpha, e\mu}(q^2)=-({\alpha\over\pi})
\int\!\int\! dx_1dx_2~[~3~\ln |1-{q^2x_1x_2\over M^2_W(x_1+x_2)}|
+{q^2(x_1+x_2-x_1x_2)\over M^2_W(x_1+x_2)-q^2x_1x_2}]
\label{E13}
\end{equation}

For a final $u\bar u$ or $d\bar d$ pair, we have to include in
the calculation of $\Gamma_{\mu,\gamma}(q^2)$ another (single $W$)
diagram as shown in Fig.\ 2b. This is, in fact, the non universal
component of such charged boson vertex, since to the graph of 
Fig.\ 2a
with two $W$s a universal quantity is associated. Strictly speaking,
diagrams with charged would-be Goldstone bosons should also be included
when the mass of the involved quarks cannot be neglected. This
corresponds to the cases of final $t\bar t$ pairs and of final
$b\bar b$ pairs. Top production will not be considered in this
paper since our technique, based on $Z$ peak subtraction, does not
apply in this case; for $b\bar b$ final states the specific
calculation of that contribution to the vertex will be given 
in Sec.\ \ref{app:cbc}. For ``light'' $u$, $d$ quarks we then obtain the
following contributions (we give the projection on the
$\widetilde{\Delta}_{\alpha, ef}(q^2, \theta)$ combination directly):

\widetext
\begin{equation}
\widetilde{\Delta}^{vert.(W)}_{\alpha, eu}(q^2)=
{\alpha\over6\pi}
\int\!\int\! dx_1dx_2~[\ln|1-{q^2x_1x_2\over M^2_W(1-x_1-x_2)}|
-{q^2(1-x_1)(1-x_2)\over M^2_W(1-x_1-x_2)-q^2x_1x_2}]
\label{E14}
\end{equation}

\begin{equation}
\widetilde{\Delta}^{vert.(W)}_{\alpha, ed}(q^2)=
2\widetilde{\Delta}^{vert.(W)}_{\alpha, eu}(q^2)
\label{E15}
\end{equation}
\noindent

\begin{figure}
\begin{center}
\epsfig{file=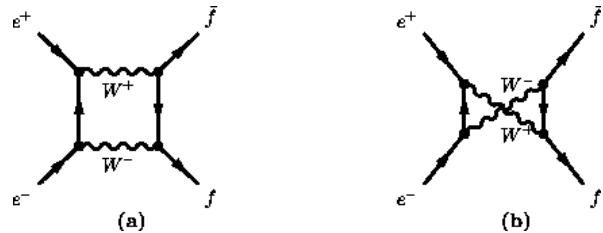,height=3truecm}
\caption{Feynman diagrams for $WW$ box contribution;
$I_{3f}=-{1\over2}$~(Fig.~3a), $I_{3f}=+{1\over2}$~(Fig.~3b).}
\end{center}
\label{fig3}
\end{figure}
The final step is now the addition of the box contributions, that
correspond to the single Feynman diagram of Fig.\ 3 ((a) for final
states with $I_{3f}=-{1\over2}$, (b) for $I_{3f}=+{1\over2}$) 
in the cases where
the mass of the final quarks can be neglected (in practice, for our
purposes, the only exception would be the box with a final $b\bar b$
pair, to be discussed in the appendix together with the related
vertex). The expression of the $WW$ box contribution to the full
scattering amplitude has been already given 
in the literature~\cite{Hollik}.
By a straightforward projection on the photon Lorentz structures one
easily obtains for final $\mu^+\mu^-$, $u\bar u$, $d\bar d$
pairs

\widetext
\begin{equation}
\widetilde{\Delta}^{Box~(WW)}_{\alpha, e\mu}(q^2, \theta)=
{\alpha\over32\pi s^4}(1-v_e)^2[I(q^2,t,M_W)+I_{5}(q^2,t,M_W)]
\label{E16}
\end{equation}

\begin{equation}
\widetilde{\Delta}^{Box~(WW)}_{\alpha, eu}(q^2, \theta)=
{3\alpha\over64\pi s^4}(1-v_e)(1-v_u)[I(q^2,u,M_W)-I_{5}(q^2,u,M_W)]
\label{E17}
\end{equation}

\begin{equation}
\widetilde{\Delta}^{Box~(WW)}_{\alpha, ed}(q^2, \theta)=
{3\alpha\over32\pi s^4}(1-v_e)(1-v_d)[I(q^2,t,M_W)+I_{5}(q^2,t,M_W)]
\label{E18}
\end{equation}

\noindent
where $v_f\equiv 1-4|Q_f|s^2$ and 
$I(q^2,t,M_W)$, $I_{5}(q^2,t,M_W)$,$I(q^2,u,M_W)$, $I_{5}(q^2,u,M_W)$
are known functions~\footnote{
We thank W. Hollik for sending us his Fortran implementation of these
functions.
}
defined in Sec.\ \ref{app:bd}, 
with $t=-{q^2\over2}(1-cos\theta)$, $u=-{q^2\over2}(1+cos\theta)$.

The full charged boson contribution is then given by the sum:

\widetext
\begin{equation}
\widetilde{\Delta}^{(charged~W)}_{\alpha, ef}(q^2, \theta)=
\widetilde{\Delta}^{s.e.~(WW)}_{\alpha, ef}(q^2)+
\widetilde{\Delta}^{vert.~(WW)}_{\alpha, ef}(q^2)+
\widetilde{\Delta}^{vert.~(W)}_{\alpha, ef}(q^2)+
\widetilde{\Delta}^{Box~(WW)}_{\alpha, ef}(q^2, \theta)
\label{E19}
\end{equation}
\noindent
in which the self-energy component is universal, the vertex with two Ws
is also universal except for the $m^2_t$ dependent 
terms for $b\bar b$
final states and the vertex with one $W$ and the box terms
depend on the final state.

The calculation of the charged boson contribution to the three
remaining gauge independent combinations $R$, $V^{\gamma Z}$,
$V^{Z\gamma}$ proceeds now in strict analogy with that illustrated for
$\widetilde{\Delta}_{\alpha,ef}$, replacing the initial photon 
with an initial $Z$. 
The calculations are straightforward and do not require a special
mention. Rather than illustrating them in detail with a long series of
formulae, we just refer to Secs.\ \ref{app:cbc} and \ref{app:bd} 
in which all the individual terms of the following sums are given:

\widetext
\begin{eqnarray}
R^{(charged~ W)}_{ef}(q^2,\theta)&=&
R^{s.e.~
(WW)}_{ef}(q^2)+R^{vert.~(WW)}_{ef}(q^2)+R^{vert.~(W)}_{ef}(q^2)\nonumber\\
&&
+R^{Box~ (WW)}_{ef}(q^2,\theta)
\label{E20}
\end{eqnarray}

\begin{eqnarray}
V^{\gamma Z~(charged~ W)}_{ef}(q^2,\theta)&=&
V^{s.e.~
(WW)}_{ef}(q^2)+V^{\gamma Z,~vert.~(WW)}_{ef}(q^2)+V^{\gamma
Z,~vert.~(W)}_{ef}(q^2)\nonumber\\
&&
+V^{\gamma Z, Box~ (WW)}_{ef}(q^2,\theta)
\label{E21}
\end{eqnarray}

\begin{eqnarray}
V^{Z\gamma~(charged~ W)}_{ef}(q^2,\theta)&=&
V^{s.e.~
(WW)}_{ef}(q^2)+V^{Z\gamma,~vert.~(WW)}_{ef}(q^2)+V^{
Z\gamma,~vert.(W)}_{ef}(q^2)\nonumber\\
&&
+V^{Z\gamma, Box~ (WW)}_{ef}(q^2,\theta)
\label{E22}
\end{eqnarray}

From the previous expressions Eqs.\ (\ref{E19})-(\ref{E22}),
we are now in a position to derive the
charged boson contribution to any observable of the process $e^+e^-\to
f\bar f$ ($f \neq e$). This is made possible by the expression that the related
differential cross section assumes in our formalism. We shall be
limited for the moment to the situation where polarized electron beams
are not considered (this might be a very interesting case in future
measurements, as stressed in a previous paper~\cite{pol}). In this
case, we shall write:

\begin{equation}
{d\sigma_{ef}\over dcos\theta}(q^2, \theta)=
{3\over8}(1+cos^2\theta)\sigma^{ef}_1+cos\theta \sigma^{ef}_2
\label{E23}
\end{equation}
where
\widetext
\begin{eqnarray}
\sigma^{ef}_1=&&N_f(q^2)({4\pi q^2\over3})\{
{\alpha^2(0)Q^2_f\over q^4 } [1+2\tilde{\Delta}_{\alpha,ef}(q^2, \theta)]
\nonumber\\
&&+2\alpha(0)|Q_f| 
[{q^2-M^2_Z\over
q^2((q^2-M^2_Z)^2+M^2_Z\Gamma^2_Z)}][{3\Gamma_e\over
M_Z}]^{1/2}[{3\Gamma_f\over N_f(M_Z^2) M_Z}]^{1/2}
{\tilde{v}_e \tilde{v}_f\over
(1+\tilde{v}^2_e)^{1/2}(1+\tilde{v}^2_f)^{1/2}}\nonumber\\
&&
\times[1 + \tilde{\Delta}_{\alpha, ef}(q^2, \theta)) - R_{ef}(q^2, \theta)
-4s_ec_e
\{{1\over \tilde{v}_e}V_{ef}^{\gamma Z}(q^2, \theta)+{|Q_f|\over \tilde{v}_f}
V_{ef}^{Z\gamma}(q^2, \theta)\}]\nonumber\\ 
&&+{[{3\Gamma_e\over
M_Z}][{3\Gamma_f\over N_f(M_Z^2) M_Z}]\over(q^2-M^2_Z)^2+M^2_Z\Gamma^2_Z}
[1-2R_{ef}(q^2, \theta)
-8s_ec_e\{{\tilde{v}_e\over1+\tilde{v}^2_e}V_{ef}^{\gamma
Z}(q^2, \theta)+{\tilde{v}_f|Q_f|\over (1+\tilde{v}^2_f)}
V_{ef}^{Z\gamma}(q^2, \theta)\}]\}\nonumber\\  
&& \label{E24}
\end{eqnarray}

\begin{eqnarray}
\sigma^{ef}_2=
&& {3N_f(q^2)\over4}({4\pi q^2\over3}) \{2\alpha(0)|Q_f|
[{(q^2-M^2_Z)\over
q^2((q^2-M^2_Z)^2+M^2_Z\Gamma^2_Z)}]
[{3\Gamma_e\over M_Z}]^{1/2}[{3\Gamma_f\over N_f(M_Z^2)
M_Z}]^{1/2}\nonumber\\
&&\times{1\over(1+\tilde{v}^2_e)^{1/2}(1+\tilde{v}^2_f)^{1/2}}
[1+\tilde{\Delta}_{\alpha,ef}(q^2, \theta)-R_{ef}(q^2, \theta)]
+{[{3\Gamma_e\over
M_Z}][{3\Gamma_f\over N_f(M_Z^2)
M_Z}]\over(q^2-M^2_Z)^2+M^2_Z\Gamma^2_Z}\nonumber\\
&&\times[{4\tilde{v}_e \tilde{v}_f\over(1+\tilde{v}^2_e)(1+\tilde{v}^2_f)}]
[1-2R_{ef}(q^2, \theta)-4s_ec_e
\{{1\over \tilde{v}_e}V_{ef}^{\gamma Z}(q^2, \theta)+{|Q_f|\over \tilde{v}_f}
V_{ef}^{Z\gamma}(q^2, \theta)\}]\} \nonumber\\ 
&&  
\label{E25}
\end{eqnarray}
\noindent
where $N_f(q^2)$ is the conventional color factor which contains 
standard QCD corrections at variable $q^2$
and the theoretical input of Eqs.\ (\ref{E24}),(\ref{E25}) contains
$\Gamma_e$, $\Gamma_f$, $\tilde{v}_e \equiv 1-4s^2_e$, 
$\tilde{v}_f\equiv 1-4|Q_f|s^2_f$ measured on
top of $Z$ resonance with an accuracy that is sufficient to prevent any
dangerous theoretical uncertainty in our predictions for future
$e^+e^-$ colliders, given their expected realistic experimental
accuracy, as exhaustively discussed in Refs.\ \cite{R7,R8}.

From the previous Eqs.\ (\ref{E19})-(\ref{E22}) it is now perfectly clear
how to compute, in a rigorously gauge-independent way, the charged $W$
contribution to various observables. This is the result of the
insertion in those equations of the charged boson contributions that we
have listed.

\begin{figure}
\begin{center}
\epsfig{file=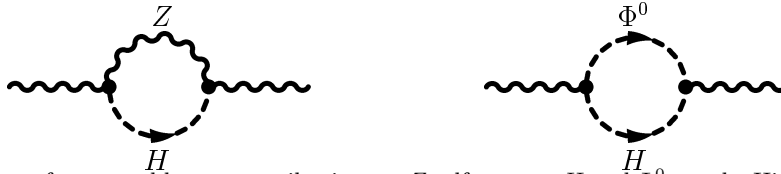,height=2.2truecm}
\caption{Feynman diagrams for neutral boson contributions to 
$Z$ self-energy. 
$H$ and $\Phi^0$ are the Higgs boson and the would-be Goldstone boson, 
respectively. }
\end{center}
\label{fig4}
\end{figure}

\begin{figure}
\begin{center}
\epsfig{file=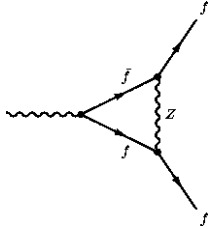,height=3truecm}
\caption{Feynman diagram for $Z$ vertex insertion.} 
\end{center}
\label{fig5}
\end{figure}

\begin{figure}
\begin{center}
\epsfig{file=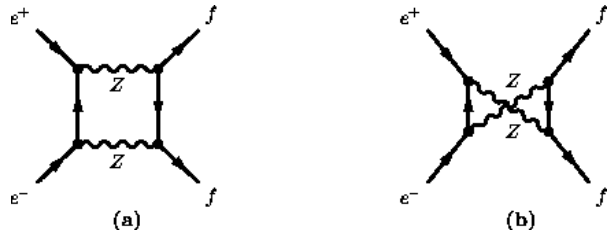,height=3truecm}
\caption{Feynman diagrams for $ZZ$ box contribution. }
\end{center}
\label{fig6}
\end{figure}

In a quite analogous way, one would compute the overall
neutral boson component or the fermionic contribution to only
self-energies, that is separately gauge independent. In fact, we have
also evaluated the neutral boson contribution by an analogous
calculation, involving $Z$ self-energies (Fig.\ 4), $Z$ vertices 
(Fig.\ 5) and $ZZ$ boxes (Fig.\ 6)
(including when necessary the neutral 
would-be Goldstone bosons and Higgs exchanges). Without
entering the details of the procedure, we anticipate that these terms
contribute, for reasonable values of the Higgs mass, all observables by
an amount that is much smaller than that coming from the charged boson
diagrams. The light fermion self-energy contribution 
(Fig.\ 7) is, on the
contrary, relevant and must be suitably taken into account for a
complete numerical estimate. This will be discussed in detail in the
two forthcoming Sections.
\begin{figure}
\begin{center}
\epsfig{file=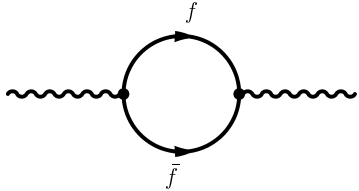,height=2.5truecm}
\caption{Feynman diagram for fermionic contribution to self-energy 
correction. }
\end{center}
\label{fig7}
\end{figure}

\section{Predictions for observables in absence of photon emission}
\label{sec:pred}

Our final step has been at this point that of translating all the
previously illustrated calculations into realistic numerical
predictions for the various observables of the process $e^+e^-\to f\bar
f$. We have concentrated our attention in an energy range $\sqrt{q^2}$
between the $Z$ resonance and 1~TeV, although we could have
considered higher energies as well. Technically speaking, we have
divided our programs into two separate phases. In the first one, we
have written analytic expressions for the four quantities
$\widetilde{\Delta}_{\alpha}$, $R$, $V^{\gamma Z}$, 
$V^{Z\gamma}$ that should contain the
bulk of the numerically relevant components (we have systematically
ignored, as one might have noticed, imaginary parts in the various
quantities; this approximation should be quite tolerable at high 
energies). More precisely, we have taken into account in a conventional way, 
i.e. including leading-log resummation, the running
of $\alpha_{QED}$, due to fermion loops, 
that appears in $\widetilde{\Delta}_{\alpha}$ (and
plays a fundamental role)~\footnote{The leptonic contribution 
to $\alpha_{QED}$ is exactly included as well as the 
top-quark effect, both below and above the $t \bar t$ threshold. The light 
quark contribution has been implemented following the generally 
adopted parameterization of Eidelman-Jegerlehner in Ref.\ \cite{alfa}.} 
and we have also inserted the full fermion
contribution to the $Z$ and $\gamma Z$ self-energies. These terms,
added to the leading terms that contain $\alpha$, $M_Z$ and the
$Z$-peak quantities, will constitute what we shall call ``non bosonic
component'' of the various observables. To these expressions we have
added those given in Eqs.\ (\ref{E19})-(\ref{E22}), 
plus the negligible neutral boson
component. These terms will generate the charged and neutral boson
contributions to all observables.

The second phase of our approach has been that of implementing our
theoretical expressions into a semi-analytical program called PALM 
(PAviaLecceMontpellier program). 
This program, which is already available, will be fully 
illustrated in a separate dedicated publication.
In PALM all the integrations 
over the Feynman variables yielding the one loop subtracted 
self-energy and vertex corrections are performed numerically~\cite{nag} 
according to semi-analytical algorithms optimized  to 
guarantee reliable and accurate results for the electroweak 
effects. In particular, the two-dimensional integrals providing the vertex 
contributions 
are first analytically reduced to one-dimensional integrals, 
that are then treated numerically on the same ground as the 
self-energy corrections. A numerical integration over the 
fermion scattering angle is performed for weak boxes in order to get 
their contribution to the integrated cross section and 
forward-backward asymmetry. It has been checked that the numerical 
procedure adopted provides results in perfect agreement with the 
formulae obtained by analytic integration over the Feynman parameters. 
The complete one loop
electroweak contributions to 
$\widetilde{\Delta}_{\alpha, ef}$, $R_{ef}$, $V^{\gamma Z}_{ef}$ and
$V^{Z \gamma}_{ef}$ are codified in PALM following a completely modular 
structure, where the corrections contributing each Lorentz structure are 
grouped into a single routine. This will allow a quite easy 
implementation in the program of analogous virtual effects, 
such those due to SUSY and other new physics models.  

To check the validity of our computation we have first compared our
numerical results, without taking QED radiation into account, 
 with the corresponding ones of the most recent version of the program 
TOPAZ0~\cite{TOPAZ0}. The input parameters used for the comparison are
$M_Z = 91.1867$~GeV, $m_{top} = 160$~GeV, $m_{Higgs} = 115$~GeV and 
$\alpha_s(M_Z) = 0.120$, together with the most recent data for $Z$ peak 
observables needed as input in PALM~\cite{bclare}. For the sake 
of comparison, the value $m_{top} = 160$~GeV has been chosen because 
fits to all electroweak data except the direct determination of $m_{top}$ 
and $M_W$ prefer a ``low'' top mass~\cite{bclare} and therefore 
this choice turns out to be consistent with 
the input $Z$ peak observables of PALM. 
Further, for a meaningful comparison with the TOPAZ0 predictions, 
the input $Z$ partial widths of PALM have been corrected for the 
effect of standard QED final-state correction. For simplicity we
have not included at this level in our program the aforementioned heavy
top contributions to the $b$ vertex and box. These actually only
contribute the $b\bar b$ differential cross section, and will affect
the total hadronic quantities that we have considered (in practice, at
this preliminary stage, the cross section for production of the five
light $u$, $d$, $s$, $c$, $b$ quarks) to a presumably negligible extent
(we shall return to this point later on).

\begin{figure}
\begin{center}
\epsfig{file=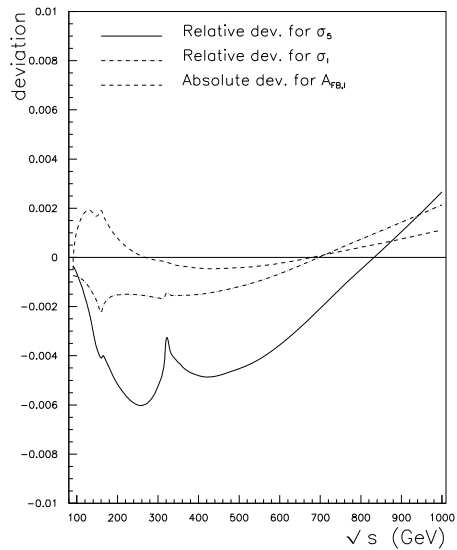,height=8.truecm}
\caption{Comparison between the predictions of PALM and 
those of TOPAZ0 for the leptonic and hadronic cross section $\sigma_l$ 
($l \neq e$) and 
$\sigma_5$ (relative deviations) and for the leptonic forward-backward 
asymmetry $A_{FB,l}$ (absolute deviation). }
\end{center}
\label{fig8}
\end{figure}
The comparison has been
carried through in an energy range $\sqrt{q^2}$ between $M_Z$ and 1~TeV. 
It is depicted graphically in Fig.\ 8 for three observables, 
chosen on the basis of their expected good experimental accuracy: the cross
section $\sigma_l$ for production of charged leptons ($l \ne e$), the related
forward-backward asymmetry $A_{FB,l}$ and the cross section for five
light quark production $\sigma_5$.

As one sees from Fig.\ 8, the agreement between our calculations
and those of TOPAZ0 is indeed impressive in the full considered energy
range. This is particularly true for the two leptonic observables, for
which the difference between the two approaches is always less than
about two permille, well below the expected experimental accuracy,
estimated to be of about a relative one percent~\cite{Zp}. This is
also true for the hadronic cross section, with the exception of two
small regions around $300$~GeV and at the end of the range when
$\sqrt{q^2}$ approaches 1~TeV, where the relative difference
reaches a five per mille value that is, though, still below the one
percent expected experimental accuracy. In fact, we believe that this
discrepancy can be originated by the heavy top effect in the $b$ vertex and
box that we did not include in our program, and are contained in
TOPAZ0. Since this difference is indeed modest, we feel
that it can be safely neglected at this stage. Note that we did not
consider the intrinsic theoretical errors implicitly contained in
both programs (e.g. PALM has as theoretical input a number of $Z$
widths and asymmetries). We estimated such uncertainties to be,
typically, of the permille size in both cases, which would still reduce
the (already miserable) existing discrepancy.

Having checked in this way the reliability of our calculation, we now
move to the real goal of this paper, that is the consideration of the
(charged) boson effects. These are shown for the three separate
variables $\sigma_{\mu}$, $A_{FB,\mu}$ and $\sigma_5$ in 
Fig.\ 9 as computed by our program PALM.

As one clearly sees from Fig.\ 9 the effect of such bosonic
contributions becomes indeed relevant as soon as the c.m. energy raises
above a typical value of approximately $200$~GeV, that corresponds to
the LEP2 limit.
At energies of about 500~GeV, in the expected range of a future
$e^+e^-$ linear collider (LC)~\cite{LC}, the relative size of the
bosonic terms is already larger than two percent in $\sigma_{\mu}$ and
six percent in $\sigma_5$. This relative effect continues to rise when
the c.m. energy increases, reaching values of the ten percent size in
the hadronic cross section in the 1~TeV limit. This clearly
generates, in our opinion, two independent and rather relevant
theoretical issues.

\begin{figure}
\begin{center}
\epsfig{file=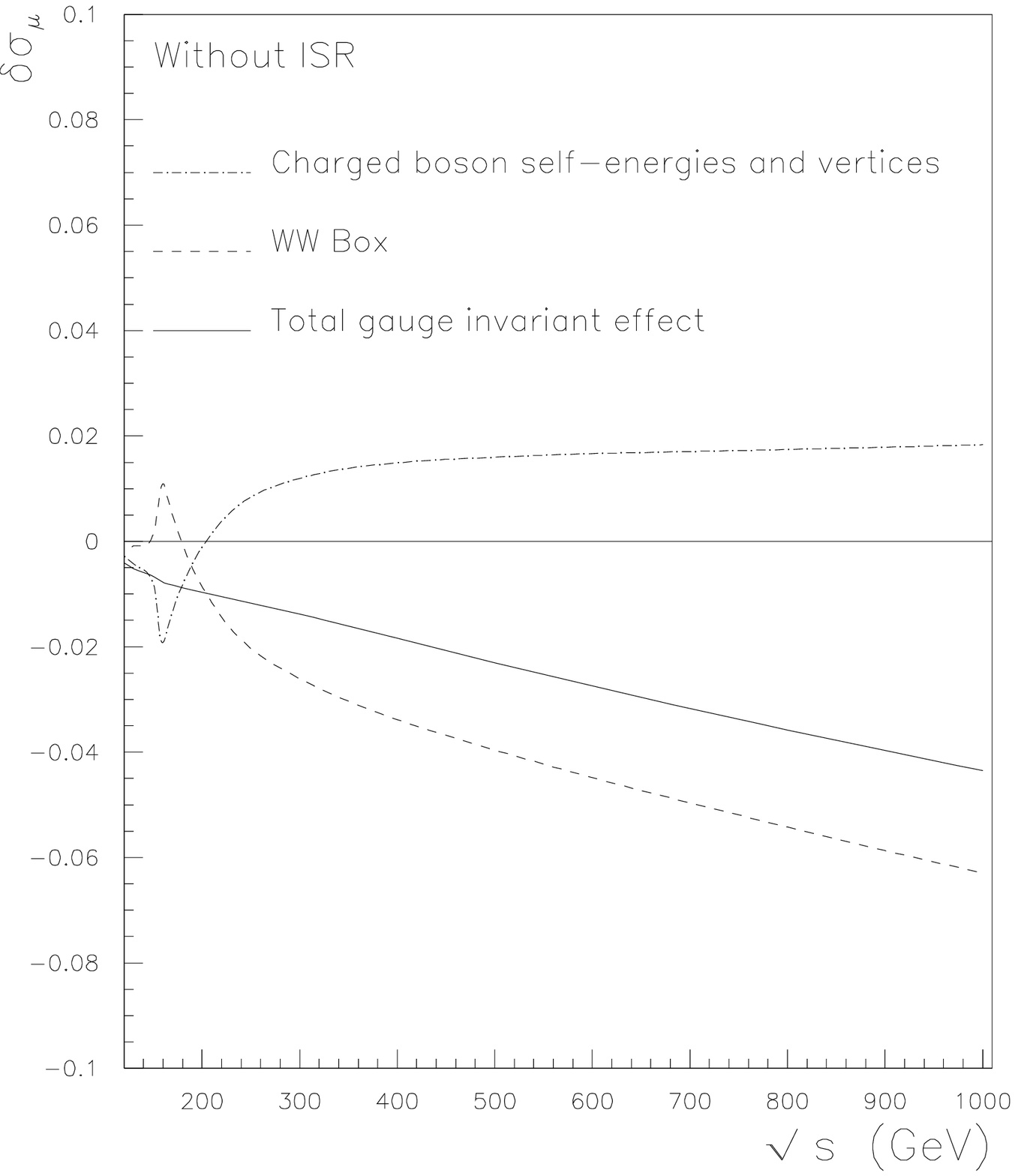,height=7.truecm}
\epsfig{file=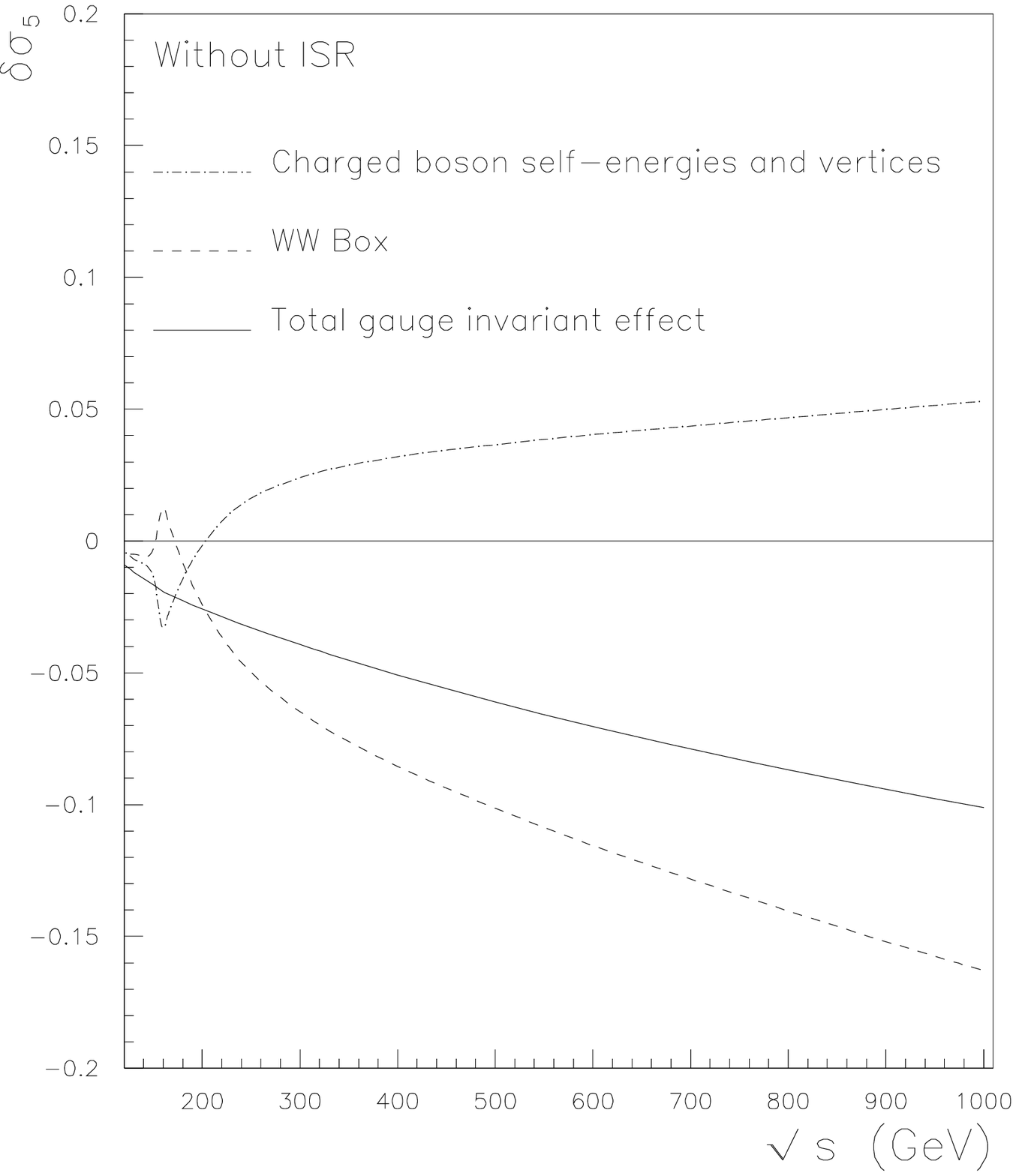,height=7.truecm}
\epsfig{file=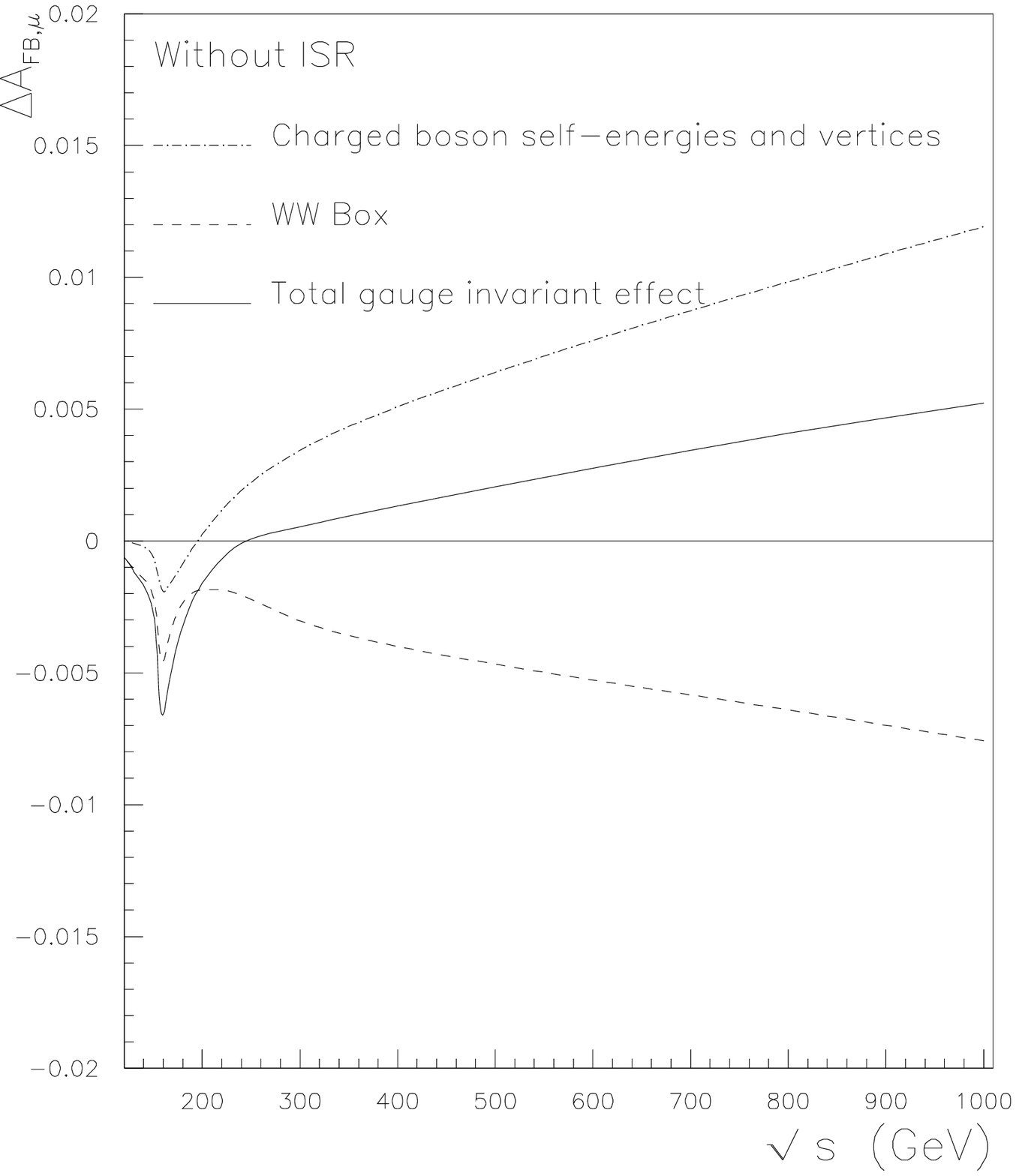,height=7.truecm}
\end{center}
\caption{The rising charged boson effects as functions of the c.m. energy, 
for the three observables $\sigma_\mu$, $\sigma_5$ and $A_{FB,\mu}$, in 
absence of photon emission. The separate effect of charged boson self-energies
and vertices (dash-dotted line), $WW$ box (dashed line) and their 
gauge invariant sum (solid line) is plotted. The relative effect of 
charged boson effects is 
shown with respect to the observables including the gauge independent 
overall neutral boson component and fermionic corrections.} 
\label{fig9}
\end{figure}

The first problem is that of understanding the reasons of such a
visible rise. At first sight, this appears to be stronger than a purely
logarithmic one, as it can be easily checked by numerical inspection.
Should this be indeed the case, one could not advocate the expected
asymptotic purely logarithmic increase dictated by renormalization
group arguments that would apply to the transverse self-energies (in
their generalized gauge independent definition proposed in
Refs.~\cite{DG1,DG2},
that also includes a small amount of ``pinched'' vertex \cite{pinch}). 
The extra
increase should therefore be rather addressed to the remaining 
component of the various combinations (by this we mean
what would remain after the purely logarithmic, separately gauge
independent, ``generalized'' Degrassi-Sirlin self-energies have been
isolated).

As a matter of fact, it has been pointed out in a previous 
reference~\cite{Moul}, and stressed again in Ref.\ \cite{DG1}, 
that the non pinch part of the $W$ vertex, Fig.\ 2b,  
behaves in the large $q^2$ limit as the squared
logarithm of $q^2/M^2_W$. Such a behaviour is often denoted as a
``Sudakov logarithm''~\cite{Sudak} since the limit 
$q^2/M^2_W\to\infty$, is
formally identical to an ``infrared'', QED like, $M_W\to 0$ limit. On
general QED analogy arguments, one would expect such terms to appear
also in the box contributions to the various combinations. More
practically, one actually sees such terms in the analytic expressions
of boxes given in Sec.\ \ref{app:bd}. Thus, the existence of such a behaviour
at large $q^2$ appears to us theoretically expected.

Strictly related with this increase is, in our opinion, the problem of
how to treat perturbatively such effects when they become too large. In
other words, one should study a mechanism of resummation analogous to
that proposed by Sudakov~\cite{Sudak} in the QED case. Although this
problem is at the moment of rather academic interest (since after all
even at 1~TeV the effect is still below ten percent), it might become
rather relevant in the case of future colliders at energies in the few
TeV range.

The second problem that arises is a consequence of the fact that the
bosonic effect is, in any case, sizeable when one moves beyond the LEP2
energy limit. This means that a proper estimate of photon emission
effects, that also includes such full bosonic terms in the quantities
to be convoluted, becomes imperative. We shall perform this calculation
in the next Sec.\ \ref{sec:qed}. This will require a technical discussion which we 
tried to make as concise as possible. 

\begin{figure}
\begin{center}
\epsfig{file=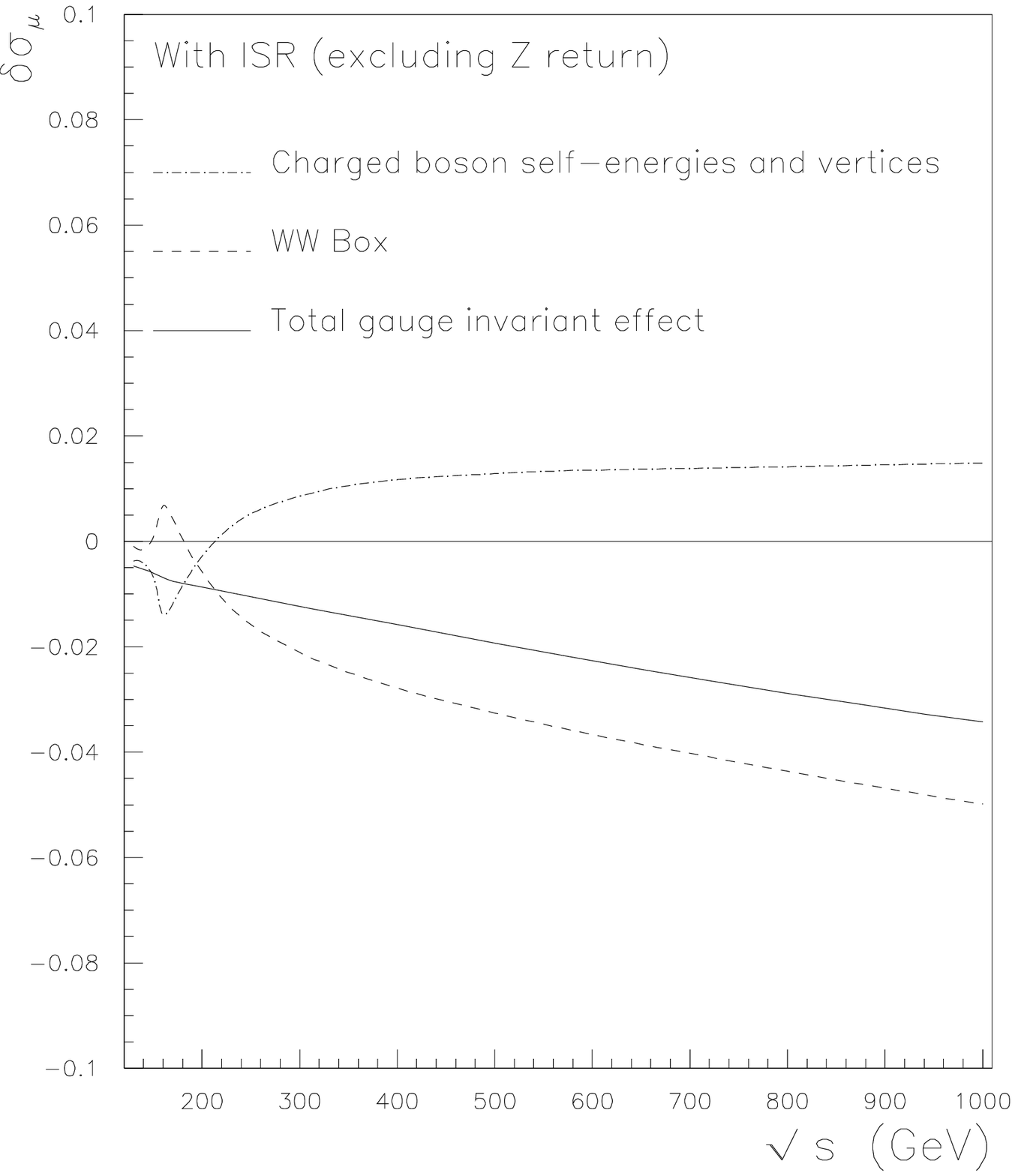,height=7.truecm}
\epsfig{file=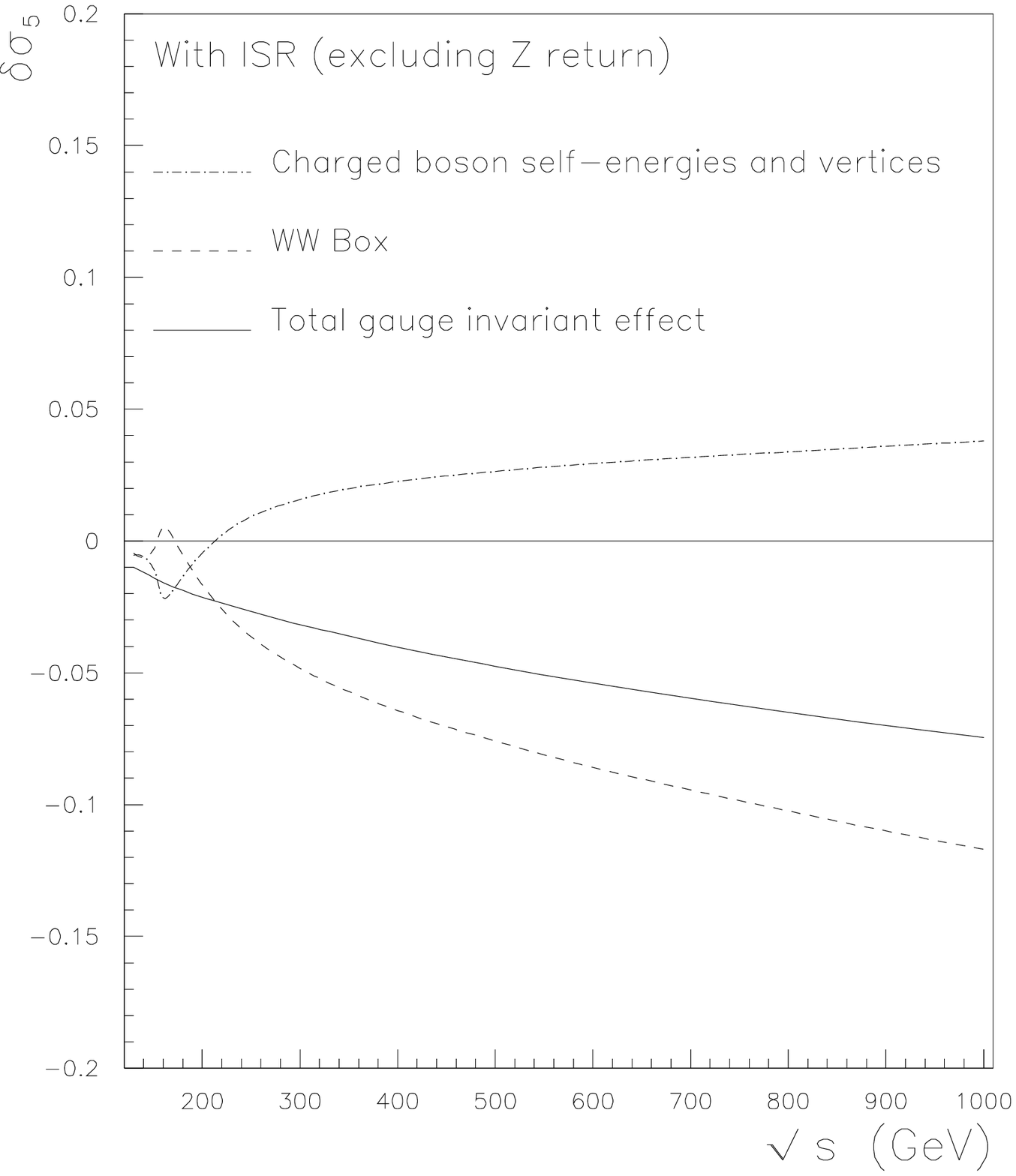,height=7.truecm}
\epsfig{file=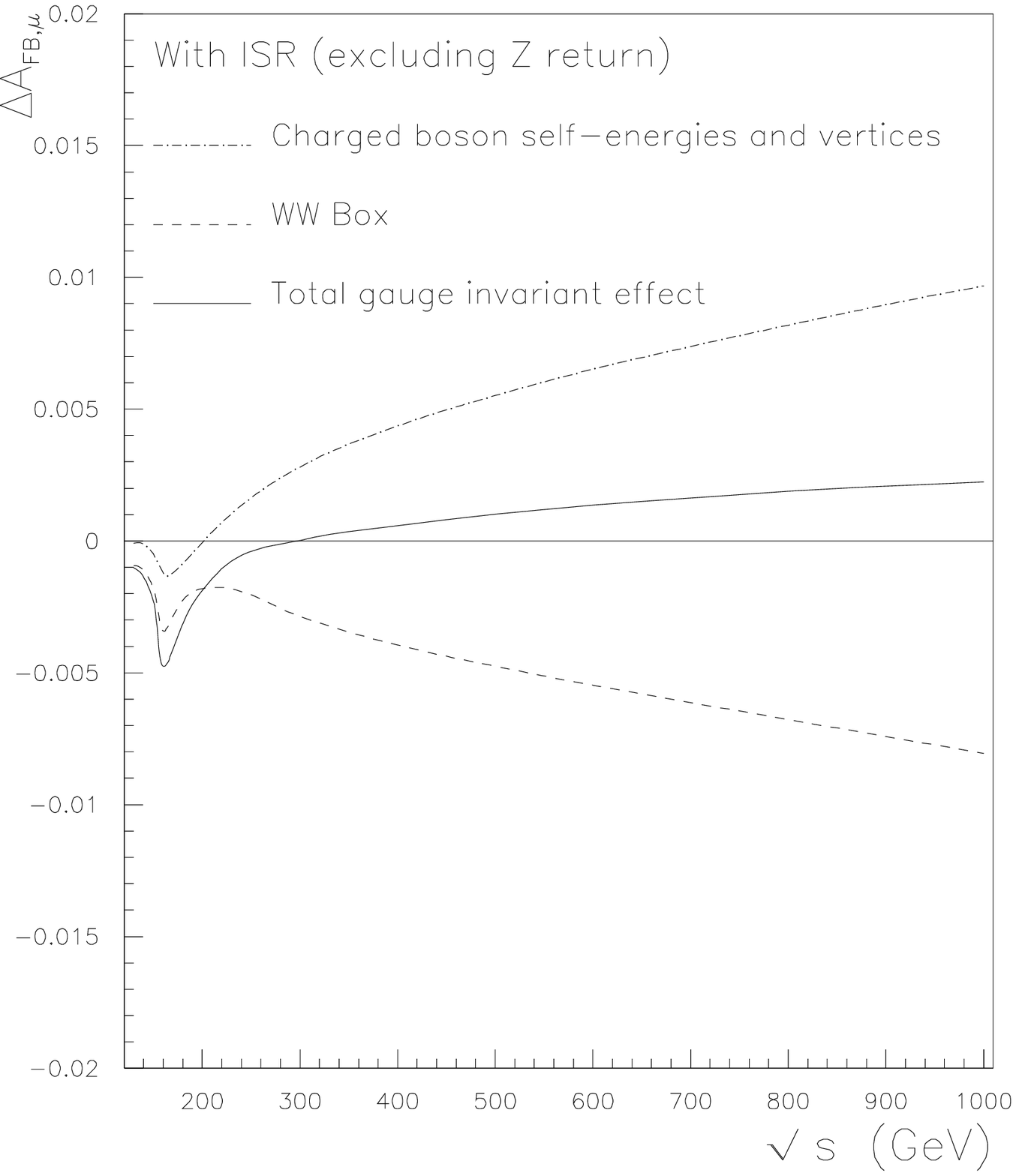,height=7.truecm}
\caption{The same as Fig.~9, including QED initial-state radiation 
(ISR). The cut $\sqrt{s'} >$ 100~GeV, $s'$ being the invariant mass of the     
event after ISR, is imposed to exclude the $Z$ radiative return.}
\end{center}
\label{fig10}
\end{figure}

\section{Treatment of photon emission effects}
\label{sec:qed}

For the aim of the present study, 
we are interested in the contribution of the initial-state radiation (ISR), 
since it is well known that it 
constitutes the largely dominating QED effect.\footnote{In principle, 
one should also 
consider other conventional QED effects as those due to final-state radiation 
and initial-final state interference. However, for the experimental 
set up considered in the following, the (very small) effect of 
final-state radiation is automatically included in our approach by means of 
the input physical quantities $\Gamma_f$. Further, the contribution of 
initial-final state interference and QED boxes has been evaluated by using the 
previously quoted program TOPAZ0 and found to be negligible as compared to 
the effect of ISR.  }
The evaluation of QED radiation effects has been 
obtained in our calculation adopting
the QED Structure Function (SF) method~\cite{conv}. It provides a simple, non 
perturbative approach, based 
on factorization theorems of infrared and collinear singularities, for 
computing leading-log (LL) photonic corrections to any arbitrary 
$e^+ e^-$ hard-scattering cross section. The method allows 
to keep under control 
the long-distance universal logarithmic contributions, due to the soft and/or 
collinear photon radiation, by convoluting a kernel cross section with 
electron(positron) structure functions. 
For a typical event selection 
where a cut on the invariant mass of the event
 after ISR is imposed, the QED corrected cross section can be 
simply calculated as~\cite{conv,prd} 
\begin{equation}
\sigma (s) = \int_0^{1 - x_{cut}} dx  \, H (x,s) 
\sigma_0 \left( (1-x) s \right)
\label{eq:conv} 
\end{equation}
where $H(x,s)$ is the radiator or flux function, and $\sigma_0$ 
is the elementary de-convoluted cross section eventually including, as 
in our case, the process-dependent 
pure weak and QCD corrections. The radiator represents the probability 
that a fraction $x$ of the whole c.m. energy $s$ is carried away by ISR 
and is calculable as:
\begin{equation}
H(x,s) = \int_{1-x}^1 {{dz} \over {z}} D (z,s) D \left( {{1-x}  \over {z}}
s \right)
\end{equation}
where $D (x,s)$ is the electron SF. A typical solution for the radiator, 
 generally used for LEP1/SLC calculations, reads~\cite{conv,prd}:
\widetext
\begin{eqnarray}
&& H(x,s) = \Delta_{2} \beta x^{\beta - 1} + h_1 (x,s) + h_2 (x,s) 
\nonumber \\ 
&& h_1 (x,s) = - {1 \over 2 } \beta (2-x)  \nonumber \\
&& h_2 (x,s) = {1 \over 8 } \beta^2 \left[ (2-x) \left( 3 \ln (1-x) 
- 4 \ln  x \right) - 4 {{\ln (1-x)} \over {x}} + x - 6 \right] \nonumber \\
&& \beta = 2 \left({\alpha \over \pi } \right) \left[ L  - 1 \right] ,
\quad L = \ln (s / m^2) , \quad 
\Delta_{2} = 1 + \left( {\alpha \over \pi } \right) \delta_{1} 
   + \left( {\alpha \over \pi } \right)^2 \delta_{2} .
\label{eq:hnt} 
\end{eqnarray}
The first exponentiated term in $H(x,s)$ describes all order 
soft multi-photon emission, $h_{1,2}(x,s)$ are associated to 
hard bremsstrahlung in collinear approximation, $\beta$ being 
the QED collinear factor. The perturbative coefficient $\Delta_{2}$ 
is the ${\cal O}(\alpha^2)$ soft+virtual $K$ factor for 
$s$-channel processes. Equation\ (\ref{eq:hnt}) for the QED radiator, 
that accurately works for precision studies of the electroweak interactions 
on top of the $Z$ resonance, becomes inadequate far from it, where, as 
discussed in Ref.\ \cite{h3}, higher-order hard photon effects 
may become numerically relevant as a consequence of the phenomenon 
of $Z$ radiative return. Actually, as pointed out in Ref.\ \cite{h3},    
${\cal O} (\alpha^2)$ next-to-leading (NL) hard-photon corrections, 
that are known from the exact ${\cal O} (\alpha^2)$ diagrammatic 
calculation~\cite{bbvn} and numerically negligible close to the $Z$ peak, 
introduce measurable effects at LEP2 and beyond and 
can be taken into account with the replacement
\begin{equation}
H (x,s) \to H (x,s) - h_2 (x,s) + \delta_2^H (1-x, s) 
\label{eq:hb}
\end{equation}
where the explicit expression for $\delta_2^H$ can be found 
in Ref.\ \cite{yrlep1}. These ${\cal O}(\alpha^2 L)$ corrections 
are of the same order of magnitude as the third-order LL ones, that 
come from the emission of three hard collinear photons 
and have been recently analyzed in the literature for their impact on 
two-fermion production cross sections at LEP2~\cite{h3}. Including 
both NL ${\cal O} (\alpha^2 )$ and LL ${\cal O} (\beta^3)$ contributions, 
an accurate QED radiator in analytic form can be cast 
as follows~\cite{h3}:
\widetext  
\begin{eqnarray}
&& H (x,s) = \Delta_{3 } \beta  x^{\beta - 1} + h_1 (x,s) 
+ \delta_2^H (1-x, s) + h_3 (x,s)  \nonumber \\
&& h_3 (x,s) = {{1} \over {3!}} \left( {\beta \over 2}\right)^3 
\left[ - {27 \over 2} + {15 \over 4 } x 
   + 4 ( 1 - {1 \over 2} x ) \left( \pi^2 - 6 \ln^2 x + 3 \hbox{\rm  Li}_2 (x) 
   \right) \right. \nonumber \\
&& \qquad \qquad + 3 \ln (1-x) \left( 7 - {6 \over x} - {3 \over  2} x \right) 
         + \ln^2 (1-x) \left( -7 + {4 \over x} + {7 \over  2} x \right) 
         \nonumber  \\
&& \qquad \qquad \left. - 6 \ln x ( 6 - x ) + 6 \ln x \ln (1-x) \left( 6 
- {4 \over x} 
   - 3 x \right) \right]  \nonumber \\
&& \Delta_{3} = \Delta_{2} + \left( {\alpha \over \pi} \right)^3 
\delta_{3} \nonumber \\
&& \delta_{3} = ( L - 1 )^3 \left( {9 \over 16} - {1 \over 2} \pi^2 
   - {4 \over 3} \psi^{(2)} (1) \right)  
\label{eq:hfull}
\end{eqnarray}
where $\psi^{(n)} (z)$ is the $n$-th order polygamma function, 
$\psi^{(n)} (z) = d^n \psi (z) / dz^n$, $\psi  (z) = \Gamma' (z) / \Gamma (z)$.
Equation\ (\ref{eq:conv}), together with the 
radiator of Eq.\ (\ref{eq:hfull}), is the formulation implemented in PALM 
for the simulation of initial-state photonic effects to the muon and hadronic 
cross section.

Since the numerator of the forward-backward asymmetry 
is a less inclusive quantity than a fully integrated cross section, 
the convolution integral for the  
asymmetry is different from that of the total cross section 
and is given by~\cite{yrlep1afb}
\widetext
\begin{equation}
A_{FB} (s) = {1 \over {\sigma(s)}} \int_{0}^{1-x_{cut}} \, dx \,
{{4 (1 - x)} \over {(2-x)^2} } \, H_{FB} (x,s) \, \sigma_{FB} ((1-x) s)
\label{eq:afbqed}
\end{equation}
where $\sigma_{FB} (s)$ is the forward minus backward 
cross section and $\sigma (s)$ is the total cross section as 
given by Eq.\ (\ref{eq:conv}). The radiator $H_{FB}(x,s)$ can be written as
\begin{equation}
H_{FB} (x,s) = H (x,s) + h_{FB,1}^{NL} (x,s) + h_{FB,2}^{LL} (x,s)
\label{eq:hfb}
\end{equation}
where $H(x,s)$ is the radiator of Eq.\ (\ref{eq:hnt}), 
$h_{FB,1}^{NL}$ stands for ${\cal O} (\alpha)$ 
NL corrections to $s$-channel processes~\cite{db91},  
$h_{FB,2}^{LL}$ are ${\cal O} (\alpha^2)$ LL ``antisymmetric'' 
corrections~\cite{bbvnring}. Equation\ (\ref{eq:afbqed}), with 
the radiator of Eq.\ (\ref{eq:hfb}), is the 
convolution integral implemented in PALM for the calculation of QED 
radiation effects to the muon asymmetry.

In the program PALM the integrations of Eqs.\ (\ref{eq:conv}) 
and (\ref{eq:afbqed}) are 
performed numerically by using a semi-analytical procedure improved by standard
importance sampling tricks to take care of the infrared and $Z$ radiative 
return peaking behaviour.

The impact of the initial-state photon emission on rising electroweak 
virtual effects is shown in Fig.\ 10 and 
Fig.\ 11, for two different typically adopted selection criteria, 
i.e. excluding the $Z$ return ($\sqrt{s'} > 100$~GeV, with $s'=(1-x)s$) 
and including it ($\sqrt{s'} > 20$~GeV). As it can be clearly
seen, a proper inclusion of QED radiation is unavoidable to get 
a reliable evaluation of bosonic virtual effects.
Actually, when excluding the $Z$ return, the QED convolution can give a 
relative reduction of the order of 25-30\% of the 
overall bosonic effects to $\sigma_{\mu}$ and $\sigma_5$ and a significant
 lowering of the same effects to $A_{FB,\mu}$. When including the $Z$ return, 
the impact of the QED convolution 
is much larger, reaching a 60-80\%  relative effect in the cross sections 
and changing sign to the net effect to the forward-backward asymmetry.
 The event selection with the inclusion of $Z$ return, in particular, is 
 responsible of a much flatter 
behaviour of the bosonic corrections, whereas the exclusion of $Z$ radiative 
return reduces the absolute size of the corrections without critically 
modifying their peculiar rising slope.
Typically, the size of the effect on the
hadronic cross section at 500~GeV is reduced by QED convolution from
six percent to four percent, which certainly represents a drastic
difference visible at an expected experimental accuracy below the $1\%$ level.

\begin{figure}
\begin{center}
\epsfig{file=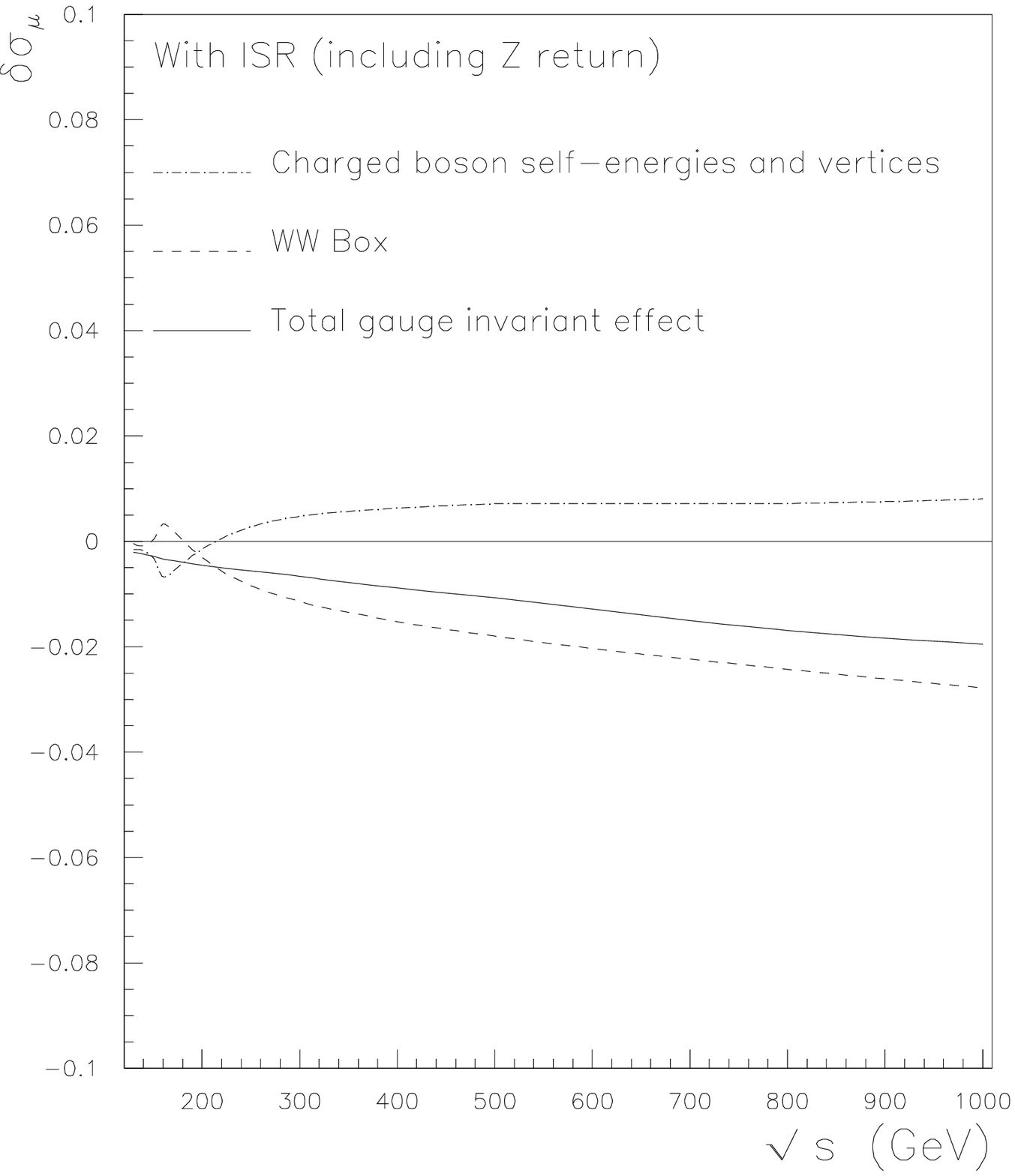,height=7.truecm}
\epsfig{file=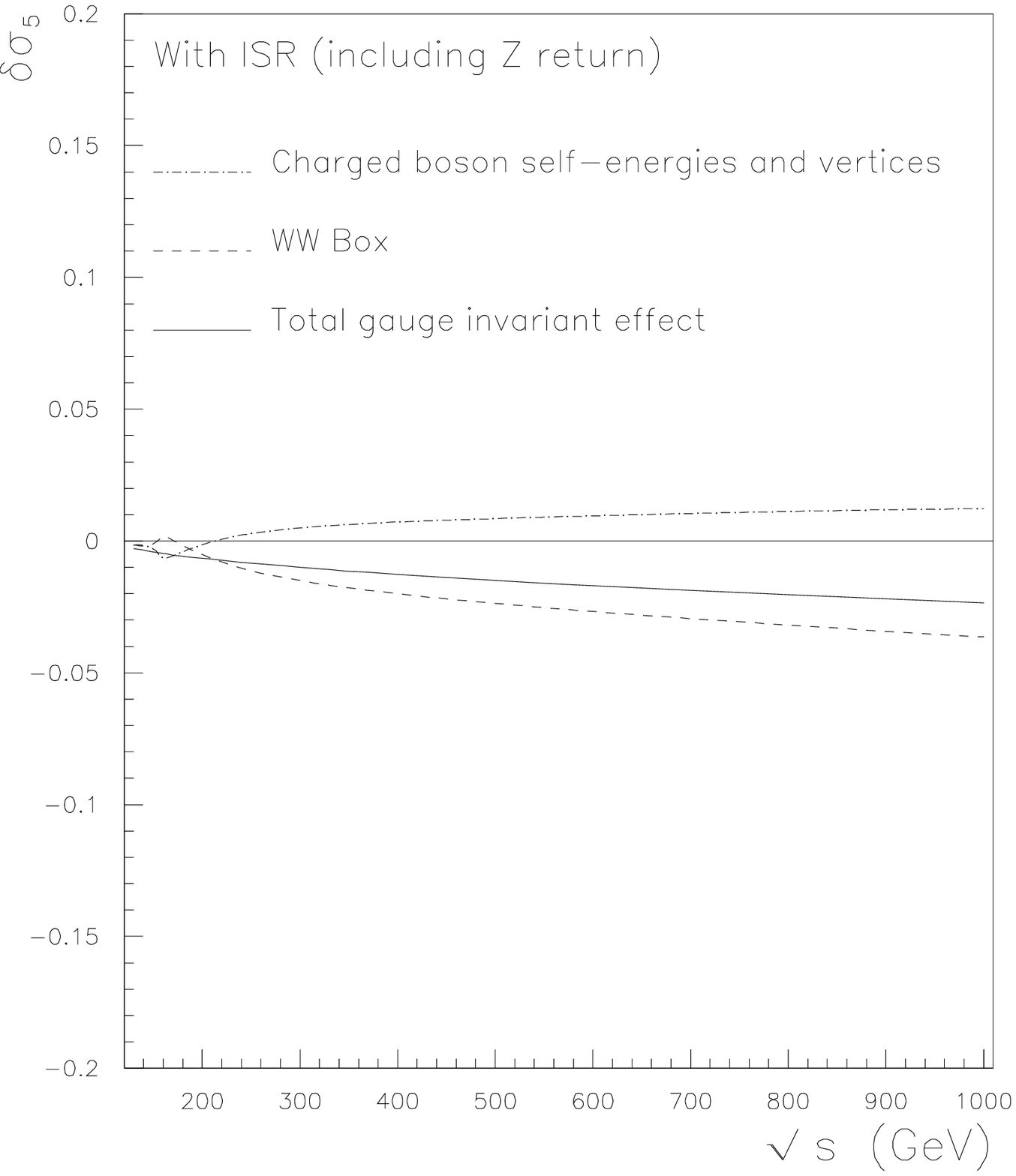,height=7.truecm}
\epsfig{file=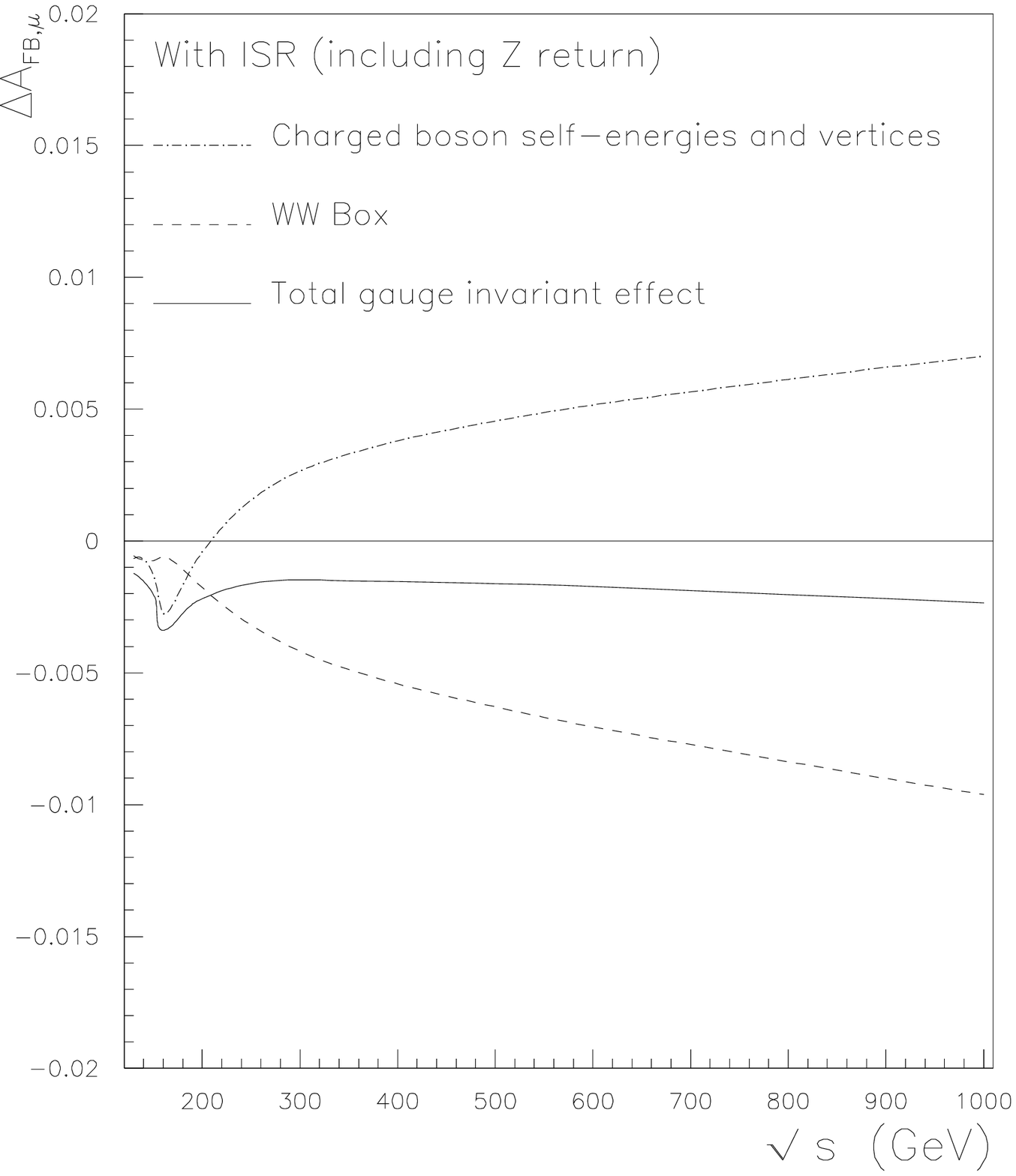,height=7.truecm}
\caption{The same as Fig.~10, with $\sqrt{s'} >$~20~GeV, to include 
the $Z$ radiative return.}
\end{center}
\label{fig11}
\end{figure}

\section{Concluding remarks}
\label{sec:concl}

We have shown in this paper that an accurate computation of theoretical
predictions for the process of $e^+e^-$ annihilation into charged
fermion pairs at future high energy colliders requires the full
determination of the bosonic contributions, including boxes
which are usually neglected on top of the $Z$ resonance. We
have performed the relevant calculations in the t'Hooft gauge, but the
\underline{overall} contributions that we have shown are completely
gauge-independent. 
For this overall effect we conclude that, since the
contribution from its ``generalized'' gauge independent self-energies can
only be of logarithmic type when $q^2$ becomes large, extra stronger
effects of Sudakov type can only be produced by the combination of the
``non pinched'' parts of vertices with the boxes. These rising effects
are evident in our calculation, and become visible as soon as one
crosses the 200~GeV range. As we have shown, the effect of a proper
inclusion of QED radiation is indeed numerically essential 
and should be carefully taken into account in the theoretical predictions.  

One possible interesting generalization of our investigation would be the
search for analogous sizeable contributions in a SUSY calculation of
virtual effects at future colliders. On a qualitative basis, we would
expect the presence of important effects produced by either
chargino or even neutralino vertices and boxes~\cite{susy}. 
Such a possibility is
at the moment under investigation.

\acknowledgements

We wish to thank Oreste Nicrosini for several fruitful discussions 
and suggestions.

\appendix
\section*{Complete 1-loop electroweak contributions to 
$\widetilde{\Delta}_{\alpha, ef}$, $R_{ef}$, $V^{\gamma Z}_{ef}$ and
$V^{Z \gamma}_{ef}$}

We give here the explicit expressions of the quantities that appear in 
Eqs.\ (\ref{E19})-(\ref{E22}).

\subsection{Charged boson contributions}
\label{app:cbc}

Universal self-energy $WW$ components (Fig.\ 1):\\

\widetext
\begin{eqnarray}
&&\widetilde{\Delta}^{s.e.~(WW)}_{\alpha, ef}(q^2)=
-({5\alpha\over4\pi})\{\int^1_0 \!dx 
~[1-{12\over5}x(1-x)]\ln|1-{q^2\over
M^2_W}x(1-x)|\nonumber\\ 
&&+{4\over15}
+({8M^2_W\over5q^2})\int^1_0\! dx ~\ln|1-{q^2\over
M^2_W}x(1-x)|\}
\label{E11B}
\end{eqnarray}

\begin{eqnarray}
&&R^{s.e.~(WW)}_{ef}(q^2)=-({\alpha\over4\pi})\int^1_0 dx
\{{2x(1-x)[20c^4+(1-2s^2)^2]-
20c^4\over4c^2s^2}[{q^2 \over q^2-M^2_Z}{\cal L}\nonumber\\
&&
+{x(1-x)M^2_Z\over
M^2_W-x(1-x)M^2_Z}]
+2M^2_W({-17+36s^2-16s^4\over4c^2s^2})[{q^2\over
q^2-M^2_Z}\bar{\cal L}-M^2_Z{\cal L}'_Z]\}
\end{eqnarray}
\begin{equation}
{\cal L}=\ln|{M^2_W-q^2x(1-x)\over M^2_W-M^2_Z x(1-x)}|
\end{equation}
\begin{equation}
\bar{\cal L}={1\over q^2}\ln|1-{q^2\over M^2_W}x(1-x)|-
{1\over M^2_Z}\ln|1-{M^2_Z\over M^2_W}x(1-x)|
\end{equation}
\begin{equation}
{\cal L}'_Z=-({1\over M^2_Z})[{x(1-x)\over M^2_W-M^2_Z x(1-x)}+{1\over M^2_Z}
\ln|1-{M^2_Z\over M^2_W}x(1-x)|]
\end{equation}

\begin{eqnarray}
&&V^{s.e.~(WW)}_{ef}(q^2)=-({\alpha\over4\pi})\int^1_0 \!dx
\{{1\over q^2}[{8s^2-9\over cs}M^2_W-5{c\over
s}q^2+{11-12s^2\over cs}x(1-x)q^2]\ln|1-{q^2\over
M^2_W}x(1-x)|\nonumber\\
&&
-({1\over M^2_Z})[{8s^2-9\over cs}M^2_W-5{c\over s}M^2_Z+
{11-12s^2\over cs}x(1-x)M^2_Z]\ln|1-{M^2_Z\over M^2_W}x(1-x)|
\}
\end{eqnarray}

$WW$ vertex component (Fig.\ 2a, universal for light fermions)\\
\widetext

\begin{equation}
\widetilde{\Delta}^{vert.~(WW)}_{\alpha, ef}(q^2)=-({\alpha\over\pi})
\int\!\int\! dx_1dx_2~[3\ln|1-{q^2x_1x_2\over M^2_W(x_1+x_2)}|
+{q^2(x_1+x_2-x_1x_2)\over M^2_W(x_1+x_2)-q^2x_1x_2}]
\label{E13B}
\end{equation}

\begin{eqnarray}
R^{vert.~(WW)}_{ef}(q^2)&=&{\alpha\over\pi}({c^2\over s^2})
\int\!\int dx_1dx_2
\{3\ln|{M^2_W(x_1+x_2)-q^2x_1x_2\over M^2_W(x_1+x_2)-M^2_Z x_1x_2}|
\nonumber\\
&&+{(q^2-M^2_Z)M^2_W(x_1+x_2)(x_1+x_2-x_1x_2)\over
[M^2_W(x_1+x_2)-q^2x_1x_2][M^2_W(x_1+x_2)-M^2_Z x_1x_2]}\}\nonumber\\
&&
\end{eqnarray}

\begin{eqnarray}
V^{vert.~(WW)}_{ef}(q^2)&=&{\alpha\over2\pi}({c\over s})
\int\!\int dx_1dx_2
\{3\ln|{M^2_W(x_1+x_2)-q^2x_1x_2\over M^2_W(x_1+x_2)-M^2_Z x_1x_2}|
\nonumber\\
&&+{(q^2-M^2_Z)M^2_W(x_1+x_2)(x_1+x_2-x_1x_2)\over
[M^2_W(x_1+x_2)-q^2x_1x_2][M^2_W(x_1+x_2)-M^2_Z x_1x_2]}\nonumber\\
&&+({q^2-M^2_Z\over q^2})[3\ln|1-{q^2x_1x_2\over M^2_W(x_1+x_2)}|
+{q^2(x_1+x_2-x_1x_2)\over M^2_W(x_1+x_2)-q^2x_1x_2}]
\}\nonumber\\
&&
\end{eqnarray}

$WW$ vertex component for $b\bar b$ final state\\
\widetext

\begin{eqnarray}
&&\widetilde{\Delta}^{vert.~(WW)}_{\alpha,eb}(q^2)=
{\alpha\over2\pi}\int\!\int dx_1 dx_2 (3+{m^2_t\over2M^2_W})
\ln|{M^2_W(x_1+x_2)+m^2_t(1-x_1-x_2)\over
M^2_W(x_1+x_2)+m^2_t(1-x_1-x_2)-q^2x_1x_2}|\nonumber\\
&& -{\alpha\over2\pi}\int\!\int dx_1 dx_2
{q^2(x_1+x_2-x_1x_2)\over
M^2_W(x_1+x_2)+m^2_t(1-x_1-x_2)-q^2x_1x_2}
+{1\over2}\widetilde{\Delta}^{vert.~(WW)}_{\alpha,e\mu}(q^2)
\end{eqnarray} 

\begin{eqnarray}
&&R^{vert.~(WW)}_{eb}(q^2)=
{\alpha\over8\pi s^2}\int\!\int dx_1 dx_2
\{[12c^2+(1-2s^2){m^2_t\over M^2_W}] \times\nonumber\\
&&
\ln|{M^2_W(x_1+x_2)+m^2_t(1-x_1-x_2)-q^2x_1x_2\over
M^2_W(x_1+x_2)+m^2_t(1-x_1-x_2)-M^2_Z x_1x_2}|+4c^2
[{q^2(x_1+x_2-x_1x_2)\over
M^2_W(x_1+x_2)+m^2_t(1-x_1-x_2)-q^2x_1x_2}
\nonumber\\
&&
-{M^2_Z(x_1+x_2-x_1x_2)\over
M^2_W(x_1+x_2)+m^2_t(1-x_1-x_2)-M^2_Z x_1x_2}]\}
+{1\over2}R^{vert.~(WW)}_{e\mu}(q^2)
\end{eqnarray}

\begin{eqnarray}
&&V^{\gamma Z,vert.~(WW)}_{eb}(q^2)=
{\alpha\over2\pi}({c\over s})
\int\!\int dx_1dx_2
\{3\ln|{M^2_W(x_1+x_2)-q^2x_1x_2\over M^2_W(x_1+x_2)-M^2_Z x_1x_2}|
\nonumber\\
&&+{(q^2-M^2_Z)M^2_W(x_1+x_2)(x_1+x_2-x_1x_2)\over
[M^2_W(x_1+x_2)-q^2x_1x_2][M^2_W(x_1+x_2)-M^2_Z x_1x_2]}\nonumber\\
&&
+{q^2-M^2_Z\over q^2}({\alpha c
\over2\pi s })\int\!\int dx_1 dx_2
[3\ln|{M^2_W(x_1+x_2)+
m^2_t(1-x_1-x_2)-q^2x_1x_2\over M^2_W(x_1+x_2)+m^2_t(1-x_1-x_2)}|
\nonumber\\
&&+{q^2(x_1+x_2-x_1x_2)
\over M^2_W(x_1+x_2)+m^2_t(1-x_1-x_2)-q^2x_1x_2}]\}
\end{eqnarray}

\begin{eqnarray}
&&V^{Z\gamma,vert.~(WW)}_{eb}(q^2)={\alpha\over2\pi}({c\over s})
\int\!\int dx_1 dx_2\nonumber\\
&& 
({q^2-M^2_Z\over q^2})[3\ln|1-{q^2x_1x_2\over M^2_W(x_1+x_2)}|
+{q^2(x_1+x_2-x_1x_2)\over M^2_W(x_1+x_2)-q^2x_1x_2}]\nonumber\\
&& 
+({\alpha\over8\pi sc})\int\!\int dx_1 dx_2 \{ 
[12c^2+(1-2s^2){m^2_t\over M^2_W}]\times\nonumber\\
&&
\ln|{M^2_W(x_1+x_2)+m^2_t(1-x_1-x_2)-q^2x_1x_2
\over M^2_W(x_1+x_2)+m^2_t(1-x_1-x_2)-M^2_Z x_1x_2}|\nonumber\\
&&
+4c^2[{q^2(x_1+x_2-x_1x_2)
\over M^2_W(x_1+x_2)+m^2_t(1-x_1-x_2)-q^2x_1x_2}
-{M^2_Z(x_1+x_2-x_1x_2)
\over M^2_W(x_1+x_2)+m^2_t(1-x_1-x_2)-M^2_Z x_1x_2}]\}\nonumber\\
&&
\end{eqnarray}

Single $W$ vertex component (Fig.\ 2b, light fermions; 
non universal)\\
\widetext

\begin{equation}
\widetilde{\Delta}^{vert.~(W)}_{\alpha, eu}(q^2, \theta)=
{\alpha\over6\pi}
\int\!\int\! dx_1dx_2~[\ln|1-{q^2x_1x_2\over M^2_W(1-x_1-x_2)}|
-{q^2(1-x_1)(1-x_2)\over M^2_W(1-x_1-x_2)-q^2x_1x_2}]
\label{EA1}
\end{equation}

\begin{equation}
\widetilde{\Delta}^{vert.(W)}_{\alpha, ed}(q^2, \theta)=
2\widetilde{\Delta}^{vert.(W)}_{\alpha, eu}(q^2, \theta)
\label{EA2}
\end{equation}

\begin{eqnarray}
R^{vert.~(W)}_{e\mu}(q^2) &=&{\alpha\over2\pi s^2}
\int\!\int dx_1dx_2
\{\ln|{M^2_W(1-x_1-x_2)-M^2_Z x_1x_2\over M^2_W(1-x_1-x_2)-q^2 x_1x_2}|
\nonumber\\
&&+{(q^2-M^2_Z)M^2_W(1-x_1-x_2)(1-x_1)(1-x_2)\over
[M^2_W(1-x_1-x_2)-q^2x_1x_2][M^2_W(1-x_1-x_2)-M^2_Z x_1x_2]}\}
\end{eqnarray}

\begin{equation}
R^{vert.~(W)}_{eu}(q^2)=
(1-{s^2\over3})R^{vert.~(W)}_{e\mu}(q^2)
\end{equation}

\begin{equation}
R^{vert.~(W)}_{ed}(q^2)=
(1-{2s^2\over3})R^{vert.~(W)}_{e\mu}(q^2)
\end{equation}

\begin{eqnarray}
V^{vert.~(W)}_{e\mu}(q^2) &=&{\alpha\over4\pi c s}
\int\!\int dx_1dx_2
\{\ln|{M^2_W(1-x_1-x_2)-M^2_Z x_1x_2\over M^2_W(1-x_1-x_2)-q^2 x_1x_2}|
\nonumber\\
&&+{(q^2-M^2_Z)M^2_W(1-x_1-x_2)(1-x_1)(1-x_2)\over
[M^2_W(1-x_1-x_2)-q^2x_1x_2][M^2_W(1-x_1-x_2)-M^2_Z x_1x_2]}\}
\end{eqnarray}

\begin{eqnarray}
&&V^{\gamma Z,vert.~(W)}_{eu}(q^2,\theta)=
V^{vert.~(W)}_{e\mu}(q^2)
-({q^2-M^2_Z\over q^2})({\alpha c\over6\pi s})\int\!\int dx_1dx_2\{
\nonumber\\
&&
\ln|1-{q^2x_1x_2\over M^2_W(1-x_1-x_2)}|-{q^2(1-x_1)(1-x_2)
\over M^2_W(1-x_1-x_2)-q^2 x_1x_2}\}
\label{EA3}
\end{eqnarray}

\begin{eqnarray}
&&V^{\gamma Z,vert.~(W)}_{ed}(q^2,\theta)=
V^{vert.~(W)}_{e\mu}(q^2)
-({q^2-M^2_Z\over q^2})({\alpha c\over3\pi s})\int\!\int dx_1dx_2\{
\nonumber\\
&&
\{\ln|1-{q^2x_1x_2\over M^2_W(1-x_1-x_2)}|-{q^2(1-x_1)(1-x_2)
\over M^2_W(1-x_1-x_2)-q^2 x_1x_2}\}
\label{EA4}
\end{eqnarray}

\begin{eqnarray}
&&V^{Z\gamma,vert.~(W)}_{ed}(q^2,\theta)=
(1-{4s^2\over3})V^{vert.~(W)}_{e\mu}(q^2)
\label{EA5}
\end{eqnarray}

\begin{eqnarray}
&&V^{Z\gamma,vert.~(W)}_{eu}(q^2,\theta)=
(1-{2s^2\over3})V^{vert.~(W)}_{e\mu}(q^2)
\label{EA6}
\end{eqnarray}

Single $W$ vertex component for $b\bar b$ final state\\
\widetext

\begin{eqnarray}
&&\widetilde{\Delta}^{vert.~(W)}_{\alpha,lb}(q^2)=
{\alpha\over3\pi}(1+{m^2_t\over2M^2_W})\int\!\int dx_1 dx_2 
\{\ln|{M^2_W(1-x_1-x_2)+m^2_t(x_1+x_2)-q^2x_1x_2\over
M^2_W(1-x_1-x_2)+m^2_t(x_1+x_2)}|\nonumber\\
&&
 + {m^2_t\over
M^2_W(1-x_1-x_2)+m^2_t(x_1+x_2)}-{m^2_t+q^2(1-x_1)(1-x_2)\over
M^2_W(1-x_1-x_2)+m^2_t(x_1+x_2)-q^2x_1x_2}\}
\end{eqnarray} 

\begin{eqnarray}
&&R^{vert.~(W)}_{eb}(q^2)=
{\alpha\over8\pi
s^2}\int\!\int dx_1 dx_2 \{[2(1-{4s^2\over3})-{4s^2 m^2_t\over3 M^2_W}]
\nonumber\\
&&[-\ln|{M^2_W(1-x_1-x_2)+m^2_t(x_1+x_2)-q^2x_1x_2\over
M^2_W(1-x_1-x_2)+m^2_t(x_1+x_2)-M^2_Z x_1x_2}|\nonumber\\
&&+{q^2(1-x_1)(1-x_2)\over
M^2_W(1-x_1-x_2)+m^2_t(x_1+x_2)-q^2x_1x_2}
-{M^2_Z(1-x_1)(1-x_2)\over
M^2_W(1-x_1-x_2)+m^2_t(x_1+x_2)-M^2_Z x_1x_2}]\nonumber\\
&&
+[(1-{4s^2\over3}){m^2_t\over M^2_W}-{8s^2\over3}]
 m^2_t
[{1\over M^2_W(1-x_1-x_2)+m^2_t(x_1+x_2)-q^2x_1x_2}\nonumber\\
&&-{1
\over M^2_W(1-x_1-x_2)+m^2_t(x_1+x_2)-M^2_Z x_1x_2}]\}
+{1\over2}R^{vert.~(W)}_{e\mu}(q^2)
\end{eqnarray}

\begin{eqnarray}
&&V^{\gamma Z,vert.~(W)}_{eb}(q^2)={\alpha\over4\pi c s}
\int\!\int dx_1dx_2
\{\ln|{M^2_W(1-x_1-x_2)-M^2_Z x_1x_2\over M^2_W(1-x_1-x_2)-q^2 x_1x_2}|
\nonumber\\
&&+{(q^2-M^2_Z)M^2_W(1-x_1-x_2)(1-x_1)(1-x_2)\over
[M^2_W(1-x_1-x_2)-q^2x_1x_2][M^2_W(1-x_1-x_2)-M^2_Z x_1x_2]}\}
\nonumber\\
&&
+({\alpha c\over3\pi s}){q^2-M^2_Z\over q^2} (1+
{m^2_t\over 2M^2_W})\int\!\int dx_1 dx_2
[-\ln|{M^2_W(1-x_1-x_2)+m^2_t(x_1+x_2)-q^2x_1x_2\over
M^2_W(1-x_1-x_2)
+m^2_t(x_1+x_2)}|\nonumber\\
&&
+{m^2_t+q^2(1-x_1)(1-x_2)
\over M^2_W(1-x_1-x_2)+m^2_t(x_1+x_2)-q^2x_1x_2}-{m^2_t
\over M^2_W(1-x_1-x_2)+m^2_t(x_1+x_2)}]
\end{eqnarray}

\begin{eqnarray}
&&V^{Z\gamma,vert.~(W)}_{eb}(q^2)=
({\alpha\over8\pi sc})\{[2(1-{4s^2\over3})-{8s^2
m^2_t\over3 M^2_W}]\int\!\int dx_1 dx_2\nonumber\\
&&
[-\ln|{M^2_W(1-x_1-x_2)+m^2_t(x_1+x_2)-q^2x_1x_2
\over M^2_W(1-x_1-x_2)
+m^2_t(x_1+x_2)-M^2_Z x_1x_2}|\nonumber\\
&& +{q^2(1-x_1)(1-x_2)\over M^2_W(1-x_1-x_2)+m^2_t(x_1+x_2)-q^2x_1x_2}
-{M^2_Z(1-x_1)(1-x_2)\over M^2_W(1-x_1-x_2)+m^2_t(x_1+x_2)-M^2_Z
x_1x_2}]\nonumber\\
&&
+[(1-{4s^2\over3}){m^2_t\over M^2_W}-{8 s^2\over3}]
\int\!\int dx_1 dx_2 m^2_t
[{1\over M^2_W(1-x_1-x_2)+m^2_t(x_1+x_2)-q^2x_1x_2}\nonumber\\
&& -{1
\over M^2_W(1-x_1-x_2)+m^2_t(x_1+x_2)-M^2_Z x_1x_2}]
\end{eqnarray}

\subsection{Neutral boson effects}
\label{app:nbe}

$ZH$ self-energy (Fig.\ 4)\\
\widetext

\begin{eqnarray}
&&R^{s.e.~(ZH)}(q^2)=({\alpha\over8\pi s^2c^2}) \int^1_0 dx~ \{
[M^2_Z(x-2)+M^2_H(1-x)][{q^2\over q^2-M^2_Z}\bar{\cal H} \nonumber\\
&&+{x(1-x)\over M^2_Z x^2+M^2_H (1-x)} +
{1\over M^2_Z}\ln|1-{M^2_Zx(1-x)\over M^2_Z x +M^2_H(1-x)}|]\nonumber\\ 
&&-x(1-x)[{q^2\over q^2-M^2_Z}{\cal H} +
{x(1-x)M^2_Z\over M^2_Z x^2+M^2_H (1-x)}]\}
\end{eqnarray}

\begin{equation}
{\cal H}\equiv \ln|{M^2_Z x +M^2_H(1-x)-q^2x(1-x)
\over M^2_Z x^2 +M^2_H(1-x)}|
\end{equation}

\begin{equation}
\bar {\cal H} \equiv {1\over q^2}\ln|1-{q^2x(1-x)\over 
M^2_Z x +M^2_H(1-x)}|
-{1\over M^2_Z}\ln|1-{M^2_Z x(1-x)\over M^2_Z x +M^2_H(1-x)}|
\end{equation}

Single $Z$ triangle (Fig.\ 5)\\
\widetext

\begin{eqnarray}
\widetilde{\Delta}^{vert.~(Z)}_{\alpha,ef}(q^2)&=&
{\alpha(2-v^2_e-v^2_f)
\over 32\pi s^2c^2}
\int\!\int dx_1 dx_2 [-\ln|1-{q^2x_1x_2\over M^2_Z(1-x_1-x_2)}|
\nonumber\\
&&+{q^2(1-x_1)(1-x_2)\over M^2_Z(1-x_1-x_2)-q^2x_1x_2}]
\end{eqnarray}

\begin{eqnarray}
R^{vert.~(Z)}_{ef}(q^2)&=& {\alpha(2+3v^2_e+3v^2_f)
\over 32\pi s^2c^2}
\int\!\int dx_1 dx_2 [-\ln|{M^2_Z (1-x_1-x_2)-q^2x_1x_2
\over M^2_Z(1-x_1-x_2-x_1x_2)}|
\nonumber\\
&&+{q^2(1-x_1)(1-x_2)\over M^2_Z(1-x_1-x_2)-q^2x_1x_2}
-{(1-x_1)(1-x_2)\over 1-x_1-x_2-x_1x_2}]
\end{eqnarray}

\begin{eqnarray}
V^{\gamma Z,vert.~(Z)}_{ef}(q^2)&=& 
({\alpha v_e(1-v_e^2)\over64\pi s^3 c^3})
\int\!\int dx_1 dx_2 [-\ln|{M^2_Z (1-x_1-x_2)-q^2x_1x_2
\over M^2_Z(1-x_1-x_2-x_1x_2)}|
\nonumber\\
&&+{q^2(1-x_1)(1-x_2)\over M^2_Z(1-x_1-x_2)-q^2x_1x_2}
-{(1-x_1)(1-x_2)\over 1-x_1-x_2-x_1x_2}]\nonumber\\
&&
-{q^2-M_Z^2\over {q^2}}({\alpha |Q_f| v_f
\over4\pi s c})\int\!\int dx_1 dx_2
 [-\ln|1-{q^2x_1x_2\over M^2_Z(1-x_1-x_2)}|
\nonumber\\
&&+{q^2(1-x_1)(1-x_2)\over M^2_Z(1-x_1-x_2)-q^2x_1x_2}]
\end{eqnarray}

\begin{eqnarray}
V^{Z\gamma,vert.~(Z)}_{ef}(q^2)&=& ({\alpha v_f(1-v^2_f)
\over64|Q_f|\pi s^3 c^3})
\int\!\int dx_1 dx_2 [-\ln|{M^2_Z (1-x_1-x_2)-q^2x_1x_2
\over M^2_Z(1-x_1-x_2-x_1x_2)}|
\nonumber\\
&&+{q^2(1-x_1)(1-x_2)\over M^2_Z(1-x_1-x_2)-q^2x_1x_2}
-{(1-x_1)(1-x_2)\over 1-x_1-x_2-x_1x_2}]\nonumber\\
&&
-{q^2-M_Z^2 \over {q^2}}({\alpha v_e\over4\pi s c})\int\!\int dx_1 dx_2
 [-\ln|1-{q^2x_1x_2\over M^2_Z(1-x_1-x_2)}|
\nonumber\\
&&+{q^2(1-x_1)(1-x_2)\over M^2_Z(1-x_1-x_2)-q^2x_1x_2}]
\end{eqnarray}

\subsection{Box diagrams}
\label{app:bd}

In the light fermionic case the $WW$ and $ZZ$ box amplitudes have
been  written in Ref.\ \cite{Hollik} in terms of $I$ and $I_5$
functions. We have also written them in the
Passarino-Veltman description~\cite{PV} in terms of  $D_i$
functions as defined in Ref.\ \cite{ABCDHag}. We have 
checked~\cite{Layssac} that both expressions agree numerically using the
package of Ref.\ \cite{Mertig}.
The extension to the $b\bar b$ case could be done very easily
by just putting $m= m_t$ instead of $m=0$ inside
$D_i(0,0,0,0,t,q^2;M,0,m,M)$.

Projecting these amplitudes on photon and $Z$ structures then give
the following results.

$WW$ box (Fig.\ 3)\\

Defining from Fig.\ 3a 

\widetext
\begin{equation}
D^W_{\mu}=D^W_d\equiv[D_{27}-{t\over2}(D_{25}-D_{24}-D_{11})]
= {[I(q^2,t,M_W)+I_{5}(q^2,t,M_W)]\over2q^2}
\end{equation}

and from Fig.\ 3b
($\bar D$ having the meaning of
changing $t$ into $u$ inside $D$)
\begin{eqnarray}
&&D^W_u\equiv[4\bar D_{27}+{q^2\over2}(\bar D_{13}+2\bar D_{26)}
+{t\over2}(\bar D_{13}+2\bar D_{25)}+{u\over2}(\bar D_{11}
+\bar D_{12}+2\bar D_{24})]\nonumber\\
&&= {[I(q^2,u,M_W)-I_{5}(q^2,u,M_W)]\over2q^2}
\end{eqnarray}

we obtain:

\begin{equation}
\widetilde{\Delta}^{(Box, WW)}_{\alpha, ef}(q^2, \theta)=
{\alpha q^2\over16\pi |Q_f|
s^4}(1-v_e)(1-v_f)D^W_f
\end{equation}

\begin{equation}
R^{(Box, WW)}_{ef}(q^2,\theta)={c^2(M^2_Z-q^2)\alpha\over
\pi s^2 }D^W_f
\end{equation}

\begin{equation}
V^{\gamma Z~(Box, WW)}_{ef}(q^2,\theta)=
{c(M^2_Z-q^2)(1-v_e)\alpha\over
4\pi s^3}D^W_f
\end{equation}

\begin{equation}
V^{ Z\gamma ~(Box, WW)}_{ef}(q^2,\theta)=
{c(M^2_Z-q^2)(1-v_f)\alpha\over
4\pi |Q_f| s^3}D^W_f
\end{equation}

$ZZ$ box (Fig.\ 6)\\

For the $ZZ$ box, we obtain
\widetext
\begin{equation}
\widetilde{\Delta}^{(Box, ZZ)}_{\alpha, ef}(q^2, \theta)=
-({\alpha q^2\over256\pi Q_f
s^4c^4})[(1-v^2_e)(1-v^2_f)D^Z_1+v_ev_f(1-v^2_e)(1-v^2_f)D^Z_2]
\end{equation}

\begin{equation}
R^{(Box, ZZ)}_{ef}(q^2,\theta)=-I_{3f}{\alpha(M_Z^2-q^2) \over8
\pi s^2c^2}[4v_ev_fD^Z_1+(1+v^2_e)(1+v^2_f)D^Z_2]
\end{equation}

\begin{equation}
V^{\gamma Z~(Box, ZZ)}_{ef}(q^2,\theta)=
-I_{3f}{\alpha(M_Z^2-q^2) \over32
\pi s^3c^3}[2v_f(1-v^2)D^Z_1+v_e(1-v^2_e)(1+v^2_f)D^Z_2]
\end{equation}

\begin{equation}
V^{ Z\gamma ~(Box, ZZ)}_{ef}(q^2,\theta)=
-({\alpha(M_Z^2-q^2) \over64
\pi Q_f s^3c^3})[2v_e(1-v^2_f)D^Z_1+v_f(1+v^2_e)(1-v^2_f)D^Z_2]
\end{equation}

with

\begin{eqnarray}
&&D^Z_1\equiv[D_{27}+{(t-u)\over4}(D_{24}-D_{25}+D_{11}]\nonumber\\
&&-[4\bar D_{27}+{q^2\over2}(\bar D_{13}+2\bar D_{26)}
+{t\over2}(\bar D_{13}+2\bar D_{25)}+{u\over2}(\bar D_{11}
+\bar D_{12}+2\bar D_{24})]\nonumber\\
&&= {[I(q^2,t,M_Z)-I(q^2,u,M_Z)]\over2q^2}
\end{eqnarray}

\begin{equation}
D^Z_2\equiv({q^2\over4})[D_{25}-D_{24}-D_{11})]
= {[I_5(q^2,t,M_Z)+I_5(q^2,u,M_Z)]\over2q^2}
\end{equation}

\subsection{Fermion pair component}
\label{app:fbc}

The contribution from Fig.\ 7 is universal and leads to ($m_f$ 
being the mass of the fermion in the loop, 
$g_{Vf}=v_f I_{3f}$, $g_{Af}=I_{3f}$):

\begin{equation}
\widetilde{\Delta}^{s.e.~(f\bar f)}_{\alpha}(q^2)= 
{\alpha Q^2_f\over\pi} \int^1_0 dx~
2x(1-x)\ln|1-{q^2\over m_f}x(1-x)|
\end{equation}

\widetext
\begin{eqnarray}
&&R^{s.e.~(f\bar f)}(q^2)=({\alpha\over4\pi s^2c^2}) \int^1_0dx~ \{
2m_f^2 g^2_{Af}[{q^2\over q^2-M^2_Z} \bar J \nonumber\\
&&+{x(1-x)\over m^2_f-M^2_Z
x(1-x)} 
+{1\over M^2_Z}\ln|1-{M^2_Z\over m_f}x(1-x)|]\nonumber\\
&&-2(g^2_{Vf}+g^2_{Af})x(1-x)[{q^2\over q^2-M^2_Z} J +
{x(1-x)M^2_Z\over m^2_f-M^2_Z
x(1-x)}]\}
\end{eqnarray}

\begin{equation}
J \equiv \ln|{m^2_f-q^2x(1-x)\over m^2_f-M^2_Z x(1-x)}|
\end{equation}

\begin{equation}
\bar J \equiv {1\over q^2}\ln|1-{q^2\over m^2_f} x(1-x)|
-{1\over M^2_Z}\ln|1-{M^2_Z\over m^2_f} x(1-x)|
\end{equation}

\begin{equation}
V^{s.e.~(f\bar f)}(q^2)=-({\alpha Q_f g_{Vf}\over2\pi sc}) \int^1_0 dx~
2x(1-x)\ln|{m^2_f-q^2x(1-x)\over m^2_f-M^2_Z x(1-x)}|
\end{equation}


\end{document}